%% file: cef_paper.tex
\documentclass[epj]{svjour}
\usepackage{graphics}
\usepackage{amsmath}
\usepackage{amssymb}

\usepackage{feynmf}

\newcommand{\s}{\sigma}
\newcommand{\e}{\epsilon}
\renewcommand{\k}{\gv{k}}
\renewcommand{\o}{\omega}
\newcommand{\gv}[1]{{\underline{#1}}}

\begin{document}

\title{Multi-orbital Anderson models and the Kondo effect: A post-NCA study}
\author{Norbert Grewe, Torben Jabben and Sebastian Schmitt%
}                     
\authorrunning{Grewe \textit{et. al.}}
\titlerunning{Multi-orbital Anderson models and the Kondo effect}
\institute{Institut f\"{u}r Festk\"{o}rperphysik, Technische Universit\"{a}t Darmstadt, 
Hochschulstr. 6, D-64289 Darmstadt, Germany}
\date{Received: date / Revised version: date}
%
\abstract{
  The low energy region of certain transition metal compounds reveals
  dramatic correlation effects between electrons, which can be studied
  by photoelectron spectroscopy. Theoretical investigations are often
  based on multi-orbital impurity models, which reveal modified
  versions of the Kondo effect. We present a systematic study of a
  multi-orbital Anderson-like model, based on a new semi-analytical
  impurity solver which goes beyond simple modifications of the well
  known NCA. We discuss one-particle excitation spectra and in
  particular the role of level positions and Coulomb-matrix elements.
  It is shown that the low-energy region as well as the overall
  features of spectra critically depend on the model parameters and
  on the quality of the approximations used. Recent photoelectron
  experiments and corresponding existing calculations are put into
  perspective. An interesting crossover scenario between different regimes 
  of ground states with characteristically different local correlations 
  is uncovered. 
  \PACS{
    {71.10.-w}{Theories and models of many-electron systems} \and
    {71.20.-b}{Electron density of states and band structure of crystalline solids}\and
    {71.27.+a}{Strongly correlated electron systems; heavy fermions}\and
    {71.55.-i}{Impurity and defect levels}
  } 
} 

\maketitle
\section{Introduction}
\label{sec:1}

The Anderson model with a spin-degenerate local level hybridized to
a broad conduction band has served in studies of the Kondo effect
\cite{gruenerMagImp74,krishnamurtyNRGSIAMI80,krishnamurtyNRGSIAMII80} 
and, combined with approximations like DMFT, also in
investigations of the behavior of lattice-periodic systems 
\cite{pruschke:dmftNCA_HM95,georges:dmft96,grenzebach:TransportPam06}. In
connection with lattice systems it plays the role of an effective
impurity which, as part of a self consisting cycle, has to be solved
for the Greens functions of local electrons by use of appropriate
impurity solvers. Among these impurity solvers are numerical
ones like QMC \cite{hirschQMC86,wernerContinousTimeQMC06} or 
NRG \cite{wilsonNRG75,petersNewNRG06}
and so called semi-analytical
ones, which are based on direct perturbation theory with respect to
hybridization and which rely on numerical procedures to solve
implicit equations for propagators formulated via skeleton-diagram
techniques \cite{keiter:PerturbJAP71,keiterMorandiResolventPerturb84,Grewe:siam83,Kuramoto:ncaII84,bickers:nca87a}. 
The former essentially exact methods in fact can properly
describe the low energy region of the model but 
are either limited to relatively small interaction values or high 
temperatures due to numerical obstacles and exhibit
deficiencies for higher excitation energies. The newly developed
semi analytical methods like ENCA, SUNCA, and CA1 
\cite{pruschkeENCA89,haule:sunca01,greweCA108}, which
improve the well known NCA \cite{Grewe:siam83,Grewe:lnca83,Kuramoto:ncaI83,Kuramoto:ncaII84} 
via inclusion of vertex corrections
with crossing diagrams, on the other hand are well suited for the whole range of
excitation energies.

Capturing electron correlations due to strong local interactions in
realistic transition metal compounds, in the framework of band
structure theory or in form of quasiparticle band structures derived
from Greens functions, is an active field of research 
\cite{held:LDADMFTReveiw06,grewe:quasipart05,heldLDADMFT08}.
Difficulties arise from the effective local building blocks needed
in the theories, which in most cases are more complicated than a
simple Anderson impurity. 

A connected problem of equal relevance
arises in the interpretation of data from photoelectron spectroscopy
(PES) in Ce-compounds, which hopefully can demonstrate the Kondo
physics via direct observation of the formation of a many-body
resonance near the Fermi level \cite{garnier:PEMSpec97,reinertASR01}. 
Here it is the influence of
crystal-field levels and spin-orbit splitting, possibly also of
higher levels on Ce with double occupancy, which complicates the
picture and greatly enhances the necessary numerical effort. Since
the resolution of inverse PES still is not sufficiently high enough
to measure directly all details needed of the DOS above the Fermi
level $\e_F$, one relies on direct PES together with extrapolations
into a limited region of thermally occupied states above $\e_F$.
Although a consistent picture of the many-body effects in some of
the Ce-compounds arises \cite{Ehm2007}, some uncertainties remain. In
particular, comprehensive and reliable information about the whole
spectral region of energies up to the excitation energies of local
states with differing occupancy and their influence on the low-energy
region is highly desirable.

Attempts of calculating one-particle excitation spectra for
multi-orbital atoms in a metallic host in the frame of direct
perturbation theory have so far been based on the NCA as a theory
designed for the case of infinite local Coulomb repulsion 
$U$ \cite{Ehm2007,vildosolaLDANCA05},
with some simple adaptations for the additional 
orbitals \cite{sakiMultiOrbNCA05,otsukiFiniteUNCA06}. 
Vertex
corrections have thus been neglected and only two possibilities for
the occupancy of the local shell are realized. Published attempts to
improve on this kind of approach have not furnished e.g.\ improved
excitation spectra yet \cite{roura-basKondoTempUNCA07}. 
Nevertheless, it is known, that even for
the simple Anderson model important modifications arise from
better approximations and more realistic values of the local
Coulomb repulsion, concerning e.g.\ the characteristic low energy
scales as well as the bulk of the more remote spectrum above the
Fermi level \cite{pruschkeENCA89}. 

A recent study \cite{greweCA108} has demonstrated that among a
few existing qualitative improvements of NCA, some of them with even
somewhat higher accuracy, the ENCA can serve as a rather reliable
approximation in a large region of excitation energies, and that the
numerical effort is limited and can be handled well. ENCA goes 
considerably beyond
NCA; it includes all vertex corrections up to second order in the
hybridization strength and   describes the region of all finite
$U$. ENCA e.g.\  correctly reproduces the
low energy scale $T_K$ with its 
dependence on model parameters~\cite{pruschkeENCA89}. 

We have generalized the ENCA to the case of a multi-orbital
model and present here a systematic study of this new ENCACF.
We will discuss some interesting features arising from variations of
shell energies and different local Coulomb-matrix elements. As it
will turn out, consistent interpretations of many features of the
spectra calculated are possible, revealing some interesting physical
trends and hopefully enabling an improved understanding of measured
spectra.

In the following section \ref{sec:2} we shortly present the model and the
construction of our ENCACF approximation and we also compare some basic
features with earlier approximations. Section \ref{sec:3} contains our main
results in form of calculated series of spectra for varying model
parameters and relevant physical interpretations. A summary and some
remarks on future developments concludes our paper in section \ref{sec:4}.


\section{A multi-orbital Anderson model and the ENCACF approximation scheme}
\label{sec:2}

When setting up realistic models for transition metal atoms with open inner shells,
one is confronted with the problem of dealing with   several free parameters which,
though in principle to be determined consistently from first principles, can usually not 
be fixed with sufficient accuracy. For definite host materials some information
from experiments and from basic calculations like band structure theory is available
and may help to define relevant regimes. In a situation, where physics is intricate and
the  validity of approximations is questionable, an approach via a somewhat restricted
 model seems  appropriate.  We therefore will study the following 
generalization of the well known
Anderson model:
\begin{align}
  \label{hamilton}
  \hat H&=\sum_{j,\s} \Bigg[
    \e_{j} \:\hat n_{j\s}+\sum_{\k}\e_{j\k} \:\hat n_{j\k\s}
    \\  
    &\phantom{\sum_{j,\s} }
    +\frac{V_j}{\sqrt{N_0}}\sum_\k 
    \left(
      \hat{c}^\dagger_{j\k\s}\hat{f}_{j\s}
      +\hat{f}^\dagger_{j\s}\hat{c}_{j\k\s} 
    \right)
  \Bigg]
  \notag  \\      \notag 
  &+\sum_j U_{jj}\: \hat{n}_{j\uparrow}\hat{n}_{j\downarrow}
  + \sum_{j<l}\sum_{\s,\s'} U_{jl}\: \hat{n}_{j\s}\hat{n}_{l\s'}
  \quad.
\end{align}
Here $\hat{c}^\dagger_{j\k\s}$ creates an electron in band $j$ 
with crystal momentum $\k$, spin $\s$ and energy $\e_{j\k}$,
$\hat{f}^\dagger_{j\s}$ creates an electron in a local level
$j$ with spin $\s$ and energy $\e_{j}$, and 
$\hat n_{j\s}=\hat{f}^\dagger_{j\s}\hat{f}_{j\s}$,
$\hat n_{j\k\s}=\hat{c}^\dagger_{j\k\s}\hat{c}_{j\k\s}$ 
are the corresponding occupation-number operators. 
The $V_j$ denote (real) matrix elements for local
hybridization  between an electron in band $j$ and one in a
local level $j$, both with the same spin. Thereby we assume, as usual,
that one-particle states can be classified according to their
crystal symmetry  with respect to the position of the impurity.
With $j$ being the relevant quantum  number, hybridization in our 
model thus respects crystal and spin symmetry separately. Insofar 
as only the local hybridization intensities
$\Delta_j(\o)=V_j^2\rho_j^{(0)}(\o)$, with 
$\rho_j^{(0)}(\o)=\frac{1}{N_0}\sum_\k\delta(\o-\e_{j\k})$ being 
the DOS of band $j$,
enter the following calculation, one may likewise speak of one single band
and its respective symmetry components around the impurity. 

From a formal point of view, as well as from the physics of the model,
conservation of $j$ is important for the proper implementation of our
diagrammatic approach, i.e.\ in the identification of irreducible pieces.
We comment in the conclusion on  how this is to be changed 
when studying lattice systems or complexes with several centers.
In general the local Coulomb matrix elements $U_{jl}$ differ, depending
on the states $j$ and $l$ involved.

Altogether the model~(\ref{hamilton}) contains a whole set of parameters,
including the functions $\rho^{(0)}_j(\o)$. Unless one is able to derive all
of these to a good accuracy with a specific situation in mind some 
reasonable assumptions have to be made in order to restrict 
the wealth of detailed information obtainable to some typically situations and
trends. We therefore assume just  two local levels $\e_1\leq\e_2$, i.e.\
$j=1,2$, without further degeneracies besides spin, and also set
the two band dispersions $\e_{1\k}=\e_{2\k}$  equal, as
well as the corresponding hybridization matrix elements $V_1=V_2$.
These are not completely unphysical assumptions; they can be realized in a highly
symmetric situation, e.g\ for $p$-states aligned along different axes of a
cubic crystal.

For the transition metals we have in mind, the $U_{jl}$ will typical be of the order 
of the bandwidth $W$, and $U_{12}$ might be somewhat smaller than $U_{11}$
and $U_{22}$. We will consider the cases $U_{11}=U_{22}=U_{12}\equiv U$ and  
$U_{11}=U_{22}=U$ with varying $U_{12}$.

Assuming a simple-cubic structure and a half-filled tight-binding band
centered around the chemical potential at $\e_\k=\mu=0$, the unperturbed 
conduction band DOS
$\rho^{(0)}(\o)$ stretches from $\o=-3$ to $\o=3$ with van-Hove kinks 
at $\o=\pm 1$. We work with a fixed Anderson 
width $\Delta\equiv\pi V^2\rho^{(0)}(0)=0.3=W/20$. Typically $\e_1=-1$
lies in the occupied band region, $U_{jl}$  exceeds $|\e_1|$,
and $\e_2$ varies  between $\e_1$ and the upper band edge.
Thus, a typical Kondo regime in the sense of the original Anderson model 
is studied.

The model (\ref{hamilton}) allows for a series of local occupation
numbers, which in increasing order cause total shell energies 
elevated by one of the Coulomb energies $U_{jl}$. We restrict ourselves mostly 
to the three lowest values $n_{loc}=0,1,2$ by properly truncating 
the system of integral equations  to be introduced below. This implies the 
following local shell-sates with the corresponding energies:

\noindent \begin{tabular}{|l|l|l|}
\hline  $n_{loc}$ & partial occupations & energy\\
\hline  0 & $n_{1\s}=n_{2\s'}=0$ & $E_0=0$\\
\hline  1 & $n_{1\s}=1$, $n_{1-\s}=n_{2\s'}=0$ & $E_{1;1}=\e_1$\\
\hline  1 & $n_{2\s}=1$, $n_{2-\s}=n_{1\s'}=0$ & $E_{1;2}=\e_2$\\
\hline  2 & $n_{1\s}=n_{1-\s}=1$, $n_{2\s'}=0$ & $E_{2;11}=2\e_1+U_{11}$\\
\hline  2 & $n_{2\s}=n_{2-\s}=1$, $n_{1\s'}=0$ & $E_{2;22}=2\e_2+U_{22}$\\
\hline  2 & $n_{1\s}=n_{2\s'}=1$ & $E_{2;12}=\e_1+\e_2+U_{12}$.\\
&  $n_{1-\s}=n_{2-\s'}=0$ & \\
   \hline
\end{tabular}
%
The unperturbed one-particle excitation energies involve
all possible differences between two of the above levels with difference 
one in the total local occupation number $n_{loc}$. This simple fact is useful to
remember when interpreting calculated spectra  in the next section. 
For a complete classification of the high energy region 
the scheme  has to be extended to $n_{loc}=3$ and $n_{loc}=4$. 
This should be kept in mind, when spectral weight has to be accounted for,
which is missing from the calculation due to the neglect of the corresponding 
propagators.


The theoretical setup of semi analytical impurity solvers is based on the early work 
of Keiter and Kimball \cite{keiter:PerturbJAP71,keiter:PerturbPRL70} and was 
brought into a form used today about 25 years ago 
\cite{Grewe:siam83,Kuramoto:ncaII84,Kuramoto:ncaI83}. It starts as "direct" time 
ordered perturbation theory in one-particle matrix elements between local and 
band states or in transfers between neighboring sites; Wick's theorem and Feynman 
diagrammatics cannot be used due to the presence of local interactions in the 
unperturbed Hamiltonian \cite{keiterMorandiResolventPerturb84,grewe:IVperturb81}. 
An alternative
though equivalent approach uses  additional, unphysical particles (''slave
bosons'') in order to stay in the  frame of Feynman perturbation theory
\cite{colemanSlaveBoson84,kroha:NCA_CMTA_rev05}. 

In either formulation
skeleton-diagram  techniques are used to perform the infinite summations, which
are necessary to  capture the essence of the many-body effects, i.e.\ the
Kondo effect as a  prototypical example, determining the ground state and
low-lying excitations. 
The implicit equations for various types of propagators of local many body
states,
formulated via skeleton diagrams, constitute a system of singular complex 
integral equations, which apart from a few special simple cases have to be solved 
numerically. Since the perturbational setup and the particular diagram 
techniques are well documented \cite{grewe:IVperturb81,Grewe:lnca83,Grewe:siam83,keiterMorandiResolventPerturb84,bickers:nca87a}, we will 
restrict our presentation to 
the new features and approximations developed for the multi-orbital model 
of equation (\ref{hamilton}). Various approximation schemes and relevant
numerical techniques have recently been investigated and compared, regarding
their quality and validity, in applications to the Anderson model and
to simple lattice models \cite{greweCA108}. We will rely on the experience
gathered there and will adopt the ENCA in a proper generalization to the multi-orbital situation.
Direct perturbation theory involves the bonus of allowing a direct 
interpretation of diagrams in terms of physical processes  \cite{grewe:IVperturb81,keiterMorandiResolventPerturb84}. 
Therefore we will present corresponding pictures in order to shortly explain the ideas
behind the ENCACF used in the following calculations. 
\begin{fmffile}{fmf_nca1}
\begin{fmfshrink}{0.8}
\fmfcmd{%
    style_def wiggly_arrow expr p =
     cdraw (wiggly p);
     cfill (arrow p)
    enddef;
    style_def dbl_wiggly_arrow expr p =
     draw_double (wiggly p) ;
     cfill (arrow p);
    enddef;
}
\begin{figure*}
   \hspace*{1cm} (a) \hspace*{1cm} \input{sig0_nca.tex}\\[6mm]
   \hspace*{1cm} (b) \hspace*{1cm} \input{vertex_enca.tex}
  \caption{Diagrammatic representation of ionic self-energies (a)
    and vertex corrections (b) for a
    single-impurity Anderson model (SIAM) in direct perturbation theory.
    Physical processes are arranged vertically along an imaginary time
    axis (broken line) which bears an energy variable z after
    Laplace-transformation. Presence of an electron in the local shell
    is indicated via a wiggly line on this time axis. Excitations of
    band electrons (straight lines) take place at hybridization vertices
    (dots on the time axis). All line crossings in these diagrams can be
    attributed as a correction to one of the
    vertices, which are drawn as triangles.}
  \label{fig:1}
\end{figure*}
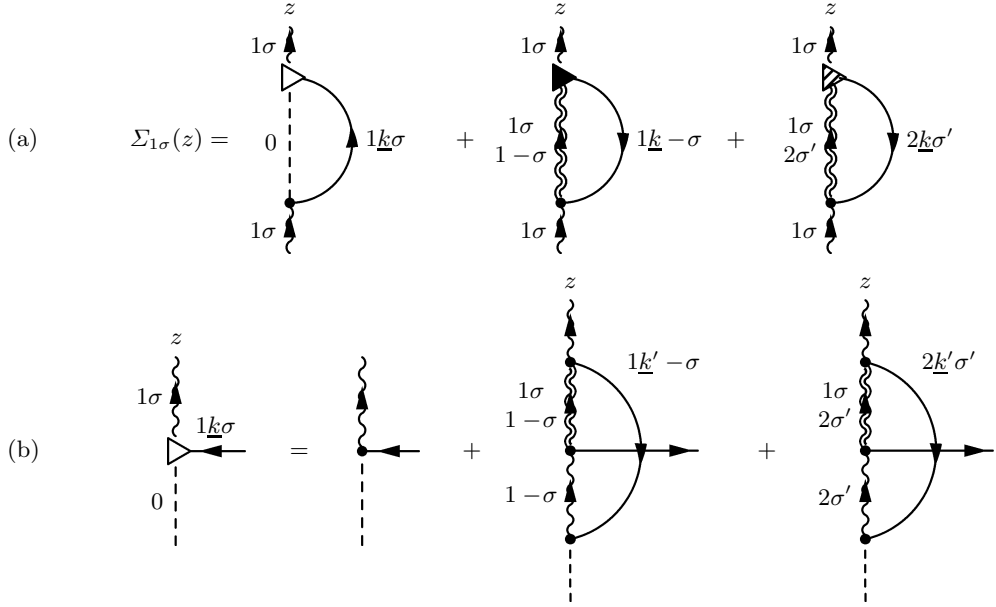
\end{fmfshrink}
\end{fmffile}
%

Propagators for states of the local shell evolve via excitation processes along
a time axis with imaginary time for technical reasons. Pieces of processes,
which are internally connected via lines visualizing excitations into the band,
constitute irreducible self-energies for the propagators. The self-energy
diagrams corresponding to a local state with just one electron of spin $\s$ 
in orbital $\e_1$ are depicted in 
Fig.~\ref{fig:1}(a).
The earliest,
i.e.\ lowest band electron line, labeled $(\k,\s')$, determines via its direction,
upward or downward, whether the following local state has zero or two electrons.
In the first diagrammatic contribution  it is the vacuum state $|0\rangle$, in the following two
cases the incoming
electron, the spin of which is preserved in the hybridization event according to
equation~(\ref{hamilton}), is absorbed either in level $\epsilon_1$ or in level $\epsilon_2$.
Two electrons in level $\epsilon_1$ 
are to have
opposite spins, i.e.\ $\s'=-\s$ here, while in the last contribution 
 the second spin-direction $\s'$
is free to
be summed over  as well as the crystal momentum $\k$. All later hybridization events
can be collected in the vertex functions, represented by the different triangles
for the three cases. Approximations to be made concern these vertex functions. In
a simple non-crossing approximation, adopted to this multi-orbital case
(SNCACF), the triangles are just replaced by bare vertices.

Although this scheme looks very simple, it is the use of full propagators for
all local lines shown, with the self-energies under debate inserted, which already
makes it highly non-trivial and allows e.g.\ for a qualitative description of
Kondo physics. Apparently, in a picture with propagators for the bare states,
series of excitations enclosing each other without crossing are generated. The
ENCA-scheme introduces the vertex corrections up to second order in the
hybridization and thus incorporates many more processes with crossing
band-electron lines. For the Anderson model it was shown to improve considerably
on the SNCA, reproducing e.g.\ the correct Kondo scale \cite{pruschkeENCA89}. 

In the generalization to our multi-orbital model the ENCACF incorporates 
vertices 
like the one shown in Fig.~\ref{fig:1}(b).

This example
suffices to clarify the building principle of the approximation. The other two vertex
functions  appearing in Fig.~\ref{fig:1}(a) are constructed in an analogous fashion. 
They depend on two energy variables, one of them 
complex, and have to be treated in all analytical equations with great care; numerically, 
their inclusion considerably enlarges the calculational effort. We will not make explicit all 
the diagrammatic self-energy equations needed and the corresponding additional set of 
diagrams necessary to build up the one-particle-Green function consistently in ENCACF. 
In principle, the diagrams are obtained from the ones of Fig.~\ref{fig:1}
and the additional 
ones not shown, by cutting one band electron line. 
Following the remarks above and the well known recipes for a 
transcription of such diagrams into formulas \cite{grewe:IVperturb81,keiterMorandiResolventPerturb84}
it is a straightforward
task to complete the systems
of integral equations for local propagators and for defect 
propagators \cite{Kuramoto:ncaIII84},
as well as to assemble all contributions to the Green functions and the 
excitation spectra derived from them.

We have solved the full ENCACF equations to convergence 
and will present one-particle excitation spectra derived from Green functions 
according to 
\begin{align}
  \label{eq:2}
  \rho_{jl\s}(\o)&=-\frac{1}{\pi}\mathrm{Im}\langle\langle\hat f_{j\s}|\hat f_{l\s}^\dagger\rangle\rangle(\hbar\o+i\delta)
  \qquad (j,l=1,2)
  \quad.
\end{align}

\begin{figure*}
  \input{figures/fig2all.tex}
  \caption{Local one-particle excitation spectrum of 
    the two-orbital model (first three curves) and of a corresponding
    Anderson model with one orbital, calculated within 
    SNCACF
    and ENCACF, respectively. The temperature $T=\frac{1}{\beta k_B}=\frac{1}{200 k_B}$
    for the two SNCACF   calculations is chosen higher than in the two   
    ENCACF cases, so that it becomes comparable with the approximate Kondo scale in
    each case. Other model parameters are $\e_1=-1$, $\e_2=0.5$,
    $\Delta=\pi\rho^{(0)}(0)=0.3$ (Anderson width) and $U_{11}=U_{22}=U_{12}=3$;
    energies are in units of the 3d=$6^{th}$ part of the width of a simple cubic
    bandstructure. 
}
  \label{fig:2}       
\end{figure*}
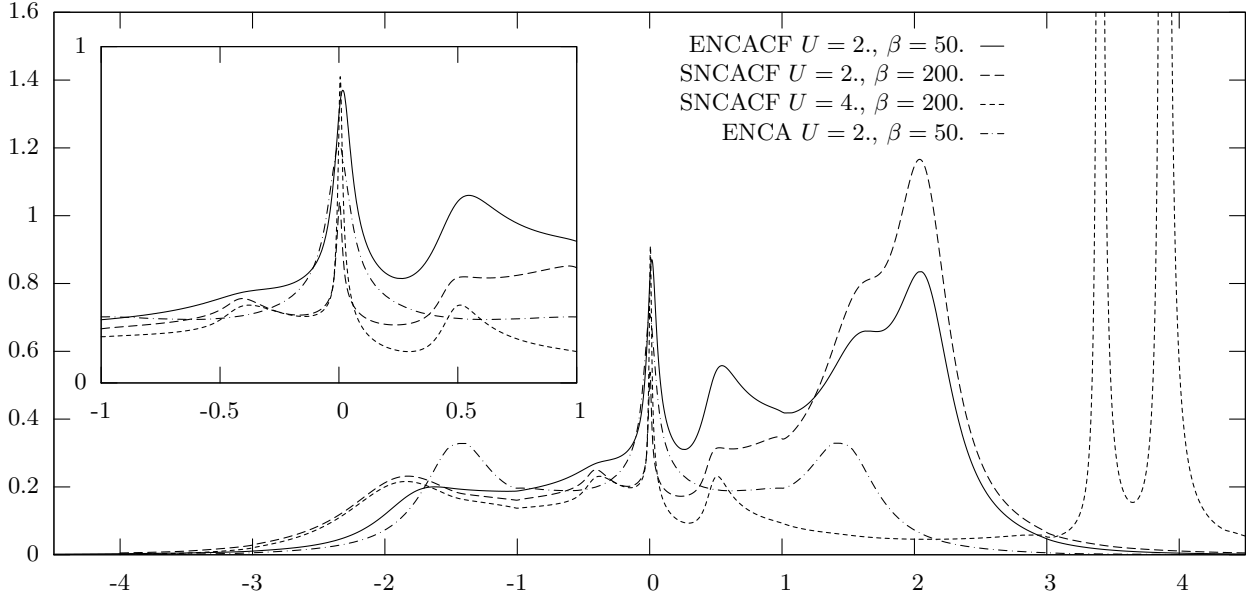
In most cases we will present the complete spectrum
\begin{align}
  \label{eq:3}
  \rho_\s(\o)&=\sum_j  \rho_{jj\s}(\o)
  \quad,
\end{align}
which in our case is normalized to 2, since it contains excitations of both local 
levels $\epsilon_1$ and $\epsilon_2$ with definite spin $\s$. First  we compare
a typical calculation with some simpler cases,
which in part are known from the literature.

In Fig.~\ref{fig:2} we present the 
a ENCACF-spectrum, a SNCACF-calculation with the same and an elevated 
value of the local repulsion, together with an ENCA-spectrum of the 
simple Anderson model without the second local level $\epsilon_2=-0.5$. 
The Coulomb-matrix elements $U_{11}=U_{22}=U_{12}$
all have the same value $U = 2$, 
except for the second SNCACF-case where $U = 4$. The inverse 
temperature is $\beta = 50$ for the ENCA-cases and $\beta = 200$ 
for 
the SNCACF, to be comparable, since SNCA underestimates the 
many-body scale considerably. Comparison of the ENCACF and 
SNCACF with the same value of $U = 2$ reveals this still to be 
true in the multi-orbital model: In fact, the spectral weight
accumulated near the chemical potential at $\o = 0$ is much larger
in the first case, although the two different temperatures 
likewise approximately reflect the respective
approximations to the  Kondo scales 
of the model without the second orbital 
(see inset of Fig. \ref{fig:2}).

A comparison with the SNCACF-spectrum at larger $U = 4$ 
demonstrates the shortcomings of a simplified calculation, 
where all doubly occupied states are projected out via letting 
U go to infinity: Considerable spectral weight is driven out 
of the band region, in form of increasingly narrow spikes. 
The ENCA-calculation without the second orbital, finally, 
helps to identify the additional spectral weight induced by 
the level $\epsilon_2$, not only in the region of high excited 
states around $\o\approx 2$, but also  near $\o = 0$, where the many-body
resonance acquires 
two satellites above and below. In a qualitative sense this behavior is known from
earlier calculations using SNCA and $U=\infty$ \cite{bickers:nca87b,kroha:CTMA97}. 
Although the number
of one-particle transitions contributing to these spectra is rather small due
to the uniform value of Coulomb-matrix elements and to the restriction to the
subspace with $n_{loc}\leq 2$, is it worthwhile to identify the bare values of transition
energies. These are for $U = 2$: 
$\Delta\e_1^{(0)}=E_{1;1}-E_0=\e_1=-1$,
$\Delta\e_2^{(0)}=E_{1;2}-E_0=\e_2=-0.5$,
$\Delta\e_3^{(0)}=E_{2;11}-E_{1;1}=\e_1+U=1$,
$\Delta\e_4^{(0)}=E_{2;12}-E_{1;1}=\e_2+U=1.5$,
$\Delta\e_5^{(0)}=E_{2;12}-E_{1;2}=\e_1+U=1$
and $\Delta\e_6^{(0)}=E_{2;22}-E_{1;2}=\e_2+U=1.5$.
Apparently in a region of energies $\o$ between $+1$ and $+2$ considerable spectral
weight is accumulated. The resonances acquire widths on the one-particle scale
$\Delta$  and in excess thereof via the local interactions and can therefore in part
not be well resolved individually. The resonance originating from  $\Delta\e_1^{(0)}$,
i.e.\ the lowest peak in the figure is known to be broadened by $2\cdot\Delta$ 
in a pure ENCA-calculation without the higher local level due to the blocking of the
resonance, if an electron with opposite spin already resides there \cite{bickers:nca87b}. This
blocking effect apparently is enhanced by the presence of the second local
level.

The visibility of excitation peaks in these spectra is largely determined by the
thermal occupation probability of the initial and final shell state. Whereas
e.g.\ the transition $\Delta\e_1^{(0)}$  possesses a weight 
$\langle\hat P_0\rangle  +\langle\hat P_{1;1\s}\rangle \approx\frac{1}{2}$
($\langle\hat P_M\rangle$ being the thermal occupation number of
the local shell-state $M$ with $\hat P_M$ the corresponding 
projection operator), the next transition at 
$\Delta\e_2^{(0)}$  has negligible weight
$\langle\hat P_0\rangle  +\langle\hat P_{1;2\s}\rangle\ll1$
in the regime considered here, with single local
occupation in level $\e_1$ being by far most probable. The high-lying peaks in the $U = 4$-case
shown in \ref{fig:2}, likewise have weight near $1$, but already indicate a crossover to narrow
spikes, since broadening begins to decrease due to a lack of nearby band states.
In this way, the rough features of single particle excitation spectra can
completely be classified.

\section{Discussion of one-particle excitation spectra and correlation effects}
\label{sec:3}
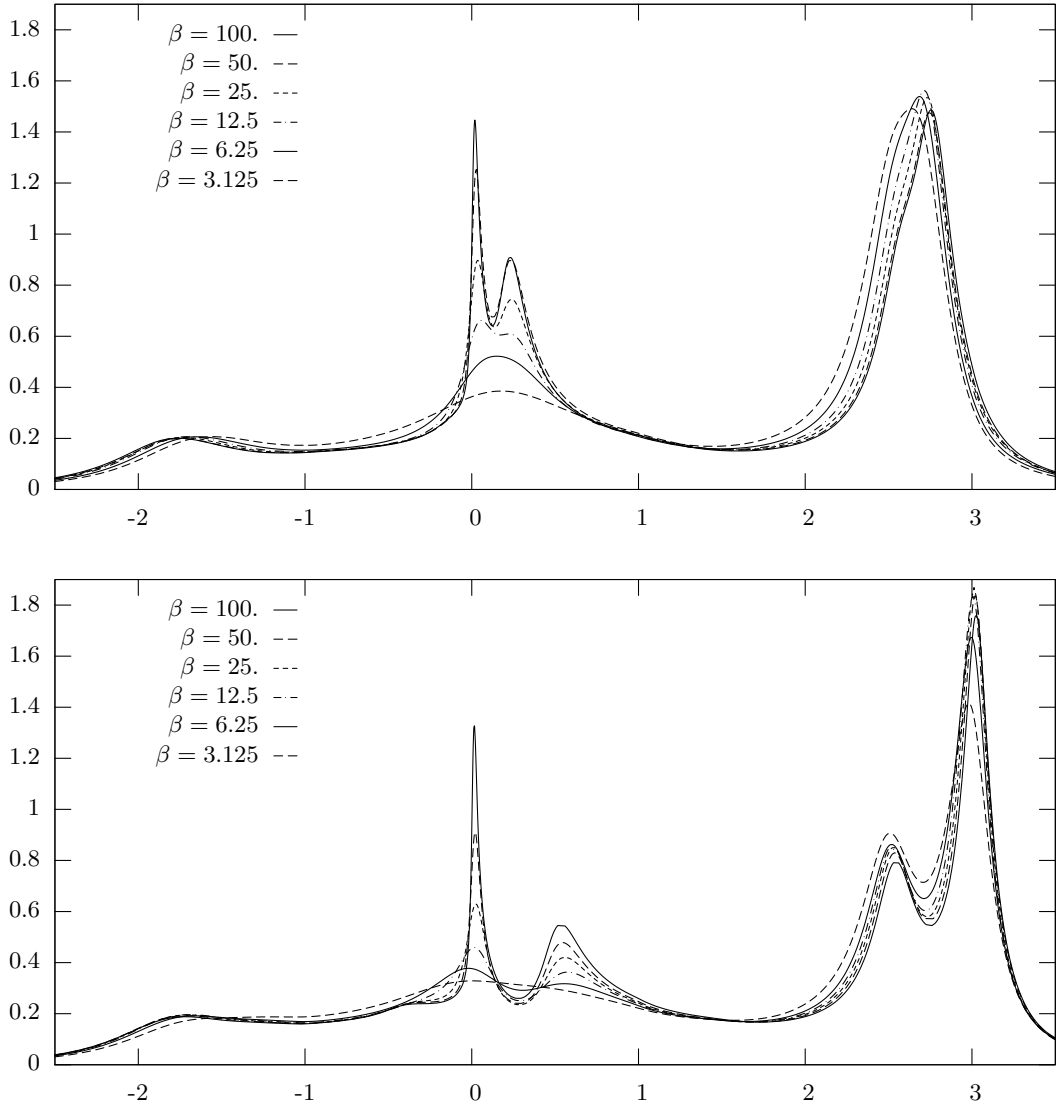
\begin{figure*}
  \input{figures/fig3a.tex}
  \input{figures/fig3b.tex}
  \caption{Temperature dependence of the local one-particle excitation spectrum
    of the two-orbital model in ENCACF-approximation. Parts (a) and (b)  differ
    in the value of the second local particle level $\e_2=-0.8$  and $\e_2=-0.5$,
    respectively. Other model parameters are as in Fig. \ref{fig:2}. 
  }
  \label{fig:3}
\end{figure*}
\begin{figure*}
  \input{figures/fig4.tex}
  \caption{Dependence of the local one-particle excitation spectrum
    of the two-orbital model on the position of the second level $\e_2$ 
    in ENCACF-approximation. The temperature $\frac{1}{50 k_B}$ compares
    roughly with the Kondo scale, other model parameters are as 
    in Fig. \ref{fig:2}. }
  \label{fig:4}
\end{figure*}
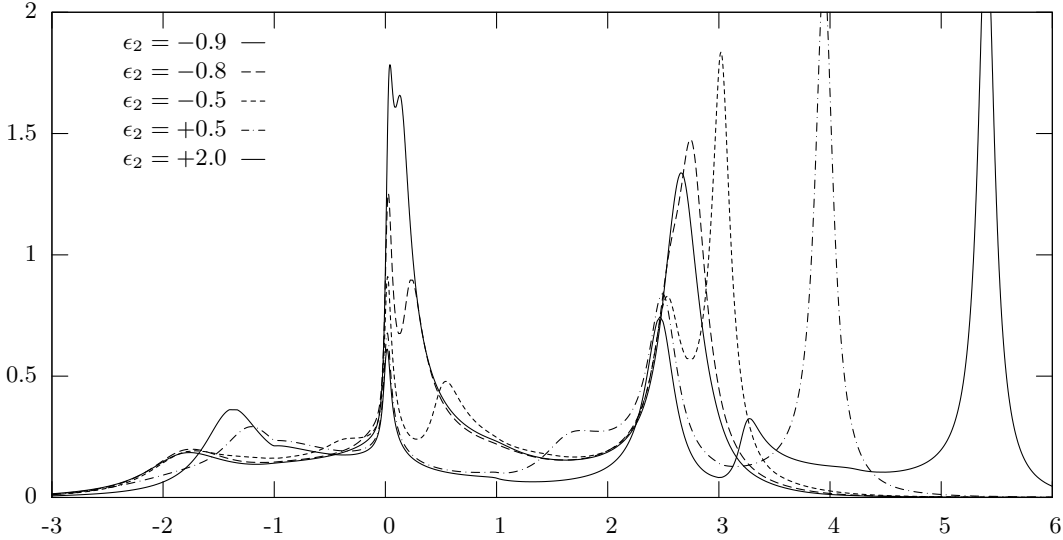

Figs.~\ref{fig:3} and ~\ref{fig:4} are designed in order to 
identify correlation effects in
the spectra, and in particular the influence of the presence
of the second local one-particle level with higher energy
$\epsilon_2>\epsilon_1$. A clear hint to the important role of
many-body correlations, beyond the interaction-induced splitting of
the local levels into a lower part at about $\epsilon_j$ and a
higher part near $\epsilon_j+U$ both with fractional weights, is a
pronounced temperature dependence in a certain interval of energies
$\omega$. 

Fig. \ref{fig:3} shows sequences of spectra for varying temperature,
typically reaching from  about a third of a characteristic low energy
scale $k_B T^*$ $(\beta=100)$ up to about ten times
$k_B T^*$. A value of about $k_B T^* \approx \frac{1}{30}$ can be
deduced from the half-width of the narrow peak at $\omega=0$; it would
coincide with the Kondo scale $k_B T_K$, if the higher local level
were not involved. Nevertheless, the choice of parameters for the
figures implies a situation like the one found in the classical
Kondo regime, with $\epsilon_2=-0.8$ or $-0.5$ clearly larger than
$\epsilon_1=-1$, both negative, but $U_{11}=U_{22}=U_{12}$ large
enough to favour a ground state occupation $n_{loc}=\sum_{j=1,2}
\sum_{\sigma} n_{j \sigma} \approx 1$. In both cases of  Fig. \ref{fig:3}, for
$\epsilon_2-\epsilon_1=0.2$ small (part a) and for
$\epsilon_2-\epsilon_1=0.5$ larger (part b), the curves in a narrow
region around the chemical potential $\omega=\mu=0$ prove to be
strongly temperature dependent, whereas the remote parts of the spectra
are much less affected by temperature variations. The qualitative
picture, known from earlier work, of a three peak many-body resonance
forming when $T^*$ is approached from above is clearly seen in
Fig. \ref{fig:3}(b). In  Fig. \ref{fig:3}(a), however, the 
lowest and weakest of these peaks is
concealed as a very flat shoulder on the left side of the main
resonance in the middle. This result again demonstrates the stronger
lifetime effects seen in the ENCACF as
compared to a simpler SNCACF-calculation, which tend to broaden the
resonances further and increase the scale $T_K$ considerably. The
origin of the three-peak structure has been explained 
consistently \cite{bickers:nca87b}: The ASR of the Kondo scenario acquires two satellites, one on each
side, due to additional local excitations in the spectral region shaped
by the many body effects. A local excitation from level $\epsilon_1$ to
level $\epsilon_2$ preserves the possibility of forming a Kondo
singlet of different kind, and the corresponding new ASR is shifted via the
foregoing excitation process by $\Delta \epsilon=\epsilon_2 -
\epsilon_1$ above the chemical potential. The inverse process leaves
its trace near $\omega=-\Delta \epsilon$; its thermal weight however,
is reduced by the small occupation probability $\epsilon_2$ compared
to the lower level $\epsilon_1$.

Fig. \ref{fig:4}  demonstrates more clearly
that this interpretation in fact holds true: The two side peaks of the
main resonance actually move with the position of the higher local
level $\epsilon_2$ and correspondingly also become less pronounced
with growing distance. The higher energy satellite can nevertheless be
followed through the high energy region around and above $\epsilon_1+U=2$;
it appears in the curve for $\epsilon_2=2$ on the right hand side of
this peak, but below the peak at energy $\epsilon_2+U=5$. The choice
of parameters for this case according to the listing of excitation
energies in the last section, only give unperturbed peak positions at
$\Delta \epsilon_{1}^{(0)}=\epsilon_1=-1 $, $\Delta
\epsilon_{2}^{(0)}=\Delta \epsilon_{3}^{(0)}=\Delta
\epsilon_{5}^{(0)}=\epsilon_2=\epsilon_1+U=2$,
$\Delta \epsilon_{4}^{(0)}=\Delta
\epsilon_{6}^{(0)}=\epsilon_2=\epsilon_2+U=5$, so that the given
identification of the feature near $\omega=3.3$ seems to be
unique. 
The temperature chosen via $\beta=50$, for  Fig. \ref{fig:3} and all the
following ones corresponds to $T^*$ or somewhat lower, so that all
many-body effects should clearly be visible. 

Also the energy region
below the chemical potentials reveals some interesting features. In
particular, the way in which the exact form of the lower main
resonance of the original Anderson model without the higher local
level is reconstructed with growing $\epsilon_2$, allows to identify
the amount of additional scattering due to the blocking between
electrons in levels $\epsilon_1$ and $\epsilon_2$. Even here, the
lower and weaker side peak can be traced for a limited region of
increasing $\epsilon_2$-values as a shallow flank, although its
thermal weight decreases and it is obscured somewhat by the changing
form and position of the main resonance arising from $\Delta
\epsilon_{1}^{(0)}=\epsilon_1=-1$. The increase of the center of this
broad main resonance with growing $\epsilon_2$ reflects the loss in
ground state energy , when level interactions become weaker. The
shrinking width of this resonance indicates the loss of blocking of
level $\epsilon_1$ due to the presence of electrons in level
$\epsilon_2$, when this energy increases.

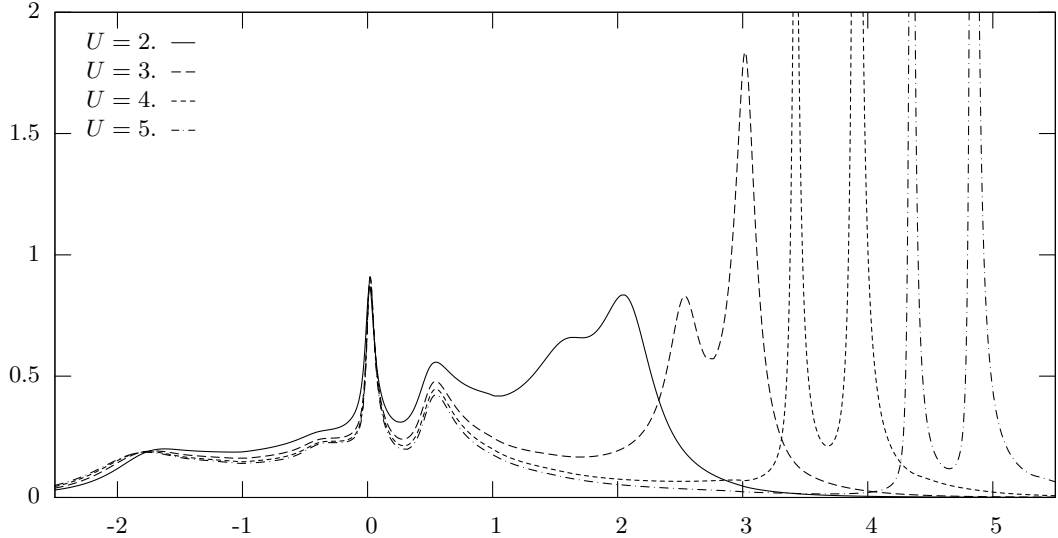
\begin{figure*}
  \input{figures/fig5.tex}
  \caption{Dependence of the local one-particle excitation spectrum
    of the two-orbital model on the Coulomb matrix element $U$ for the 
    case $U_{11}=U_{22}=U_{12}$ in ENCACF-approximation. The
    temperature and second level position are  $\frac{1}{50 k_B}$ 
    and $\e_2=-0.5$, respectively. Other model parameters are as 
    in Fig. \ref{fig:2}. 
  }
  \label{fig:5}
\end{figure*}
The next figures are devoted to a study of the influence, which the
magnitude of local Coulomb-matrix elements has on the spectra and on
the correlations. In  Fig. \ref{fig:5} we keep the equality 
$U_{11}=U_{22}=U_{12}\equiv U$, but vary $U$ in integer steps from $2$ to $5$. Unlike the
variation of temperature in  Fig. \ref{fig:3} or the variation of the upper level
position in  Fig. \ref{fig:4} the variation of U does affect the width of the
central many-body resonance. This $T^*$ gets smaller with increasing
$U$ in accord with the behavior of $T_K$ in the Kondo effect. The
decrease of $T^*$ is moderate and effectively stops when the upper
resonances $\epsilon_1 +U$ and $\epsilon_2 +U$ cross the upper band
edge at $\omega=3$. The splitting of these two highest peaks is fixed
approximately at our value of $\epsilon_2-\epsilon_1=0.5$, with
$\epsilon_1=-1$. The width of both peaks
decreases from $U=3$, in which case they just stay inside the band
region, to $U=5$, where both are situated above it. Increased scattering
inside the region with high density of band states washes these peaks
out which best can be seen in the region $1<\omega<2.5$ for the lowest
value $U=2$. The height of the central many-body resonance changes
only slightly, showing that we stay in the near saturated region
$T\lesssim T^*(U)$.

\begin{figure*}
  \input{figures/fig6.tex}
  \caption{Dependence of the local one-particle excitation spectrum
    of the two-orbital model on the particular size of $U_{12}$ of the 
    Coulomb-repulsion between an electron in level $\e_1$ and one in level 
    $\e_2$ in ENCACF-approximation. Other model parameters are as 
    in Fig. \ref{fig:2}. 
  }
  \label{fig:6}
\end{figure*}
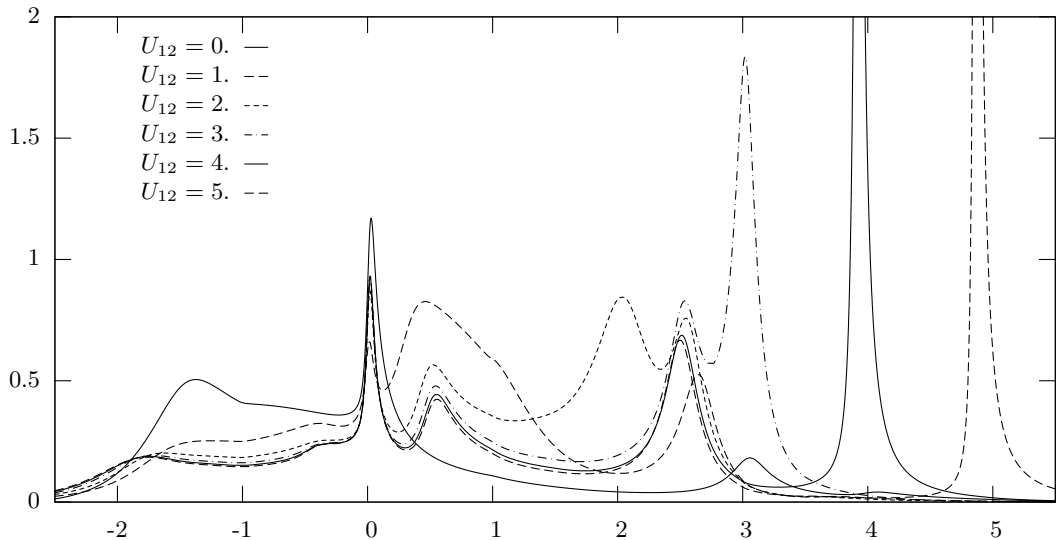
A very interesting perspective opens when variations of the
Coulomb matrix element $U_{12}$, i.e.\ the repulsion between electrons
in different local levels $\epsilon_1$ and $\epsilon_2$, are performed
keeping $U_{11}=U_{22}$ fixed.  Fig. \ref{fig:6} shows the corresponding spectra,
again for fixed $\epsilon_1=-1$, $\epsilon_2=-0.5$, $\beta=50$ and
$U_{11}=U_{22}=3$, and for a variation of $U_{12}$ in integer steps
between $0$ and $5$. It is apparent, that a considerable transfer of
spectral weight from above to below the chemical potential
takes place when going from values $U_{12}\geq 1$ to values
$U_{12}\approx 0$. In fact, the condition $\epsilon_2+U_{12}=0$,
i.e.\ $U_{12}=0.5$ marks a crossover here: For $U_{12}>0.5$ a
ground state with single level occupation $n_{loc}=1$ is stable in
zeroth order as in all cases before, whereas for $U_{12}<0.5$ the
stability goes over to a new ground state with $n_{1 \sigma}=n_{2
  \sigma}=\frac{1}{2} $, i.e.\ $n_{loc}= \sum_{j \sigma} 
n_{j \sigma}=2.$ This changes drastically the thermal occupation
probabilities of local shell states and the corresponding transitions
as well as the nature of the many-body  correlations. 

Generally, the
thermal weight of a one-particle transition  from shell state $M$ to
$M'$ is $\langle\hat{P}_M\rangle+\langle\hat{P}_{M'}\rangle $, 
i.e.\ the sum of the occupation
probabilities if the initial and the final state, in zeroth order. The
corresponding excitation energy is $\Delta \epsilon=E_{M'}-E_{M}$ and
$n^{'}_{loc}=n_{loc}+1$ by definition. For large $U_{12}$ and
$\epsilon_1<\epsilon_2$ in the regime considered here, 
$\langle\hat{P}_{1;1\sigma}\rangle$
is approximately $\frac{1}{2}$ and all other $\langle\hat{P}_M\rangle$ are
small. Consequently, the transitions with energies $\Delta
\epsilon_{1}^{(0)}=E_{1;1}-E_0=\epsilon_1$,$\Delta
\epsilon_{3}^{(0)}=E_{2;11}-E_{1;1}=\epsilon_1+U_{11}$ and $\Delta
\epsilon_{4}^{(0)}=E_{2;12}-E_{1;1}=\epsilon_2+U_{12}$ are strongly and
the others are weakly represented in the spectra. For i.e.\ $U_{12}=4$
this explains the $3$-peak structure in the high energy region with
peaks shifted somewhat outward from $\Delta \epsilon_{1}^{(0)}=-1$,
$\Delta\epsilon_{3}^{(0)}=2$, and $\Delta\epsilon_{4}^{(0)}=3.5$. For small $U_{12}$,
on the other hand , $\langle\hat{P}_{2;1\sigma2\sigma'}\rangle$ is approximately
$\frac{1}{4}$ and the other probabilities are small, so that only the
transitions with energy $\Delta\epsilon_{4}^{(0)}$
and $\Delta\epsilon_{5}^{(0)}=E_{2;12}-E_{1;2}$ are strong and the others
weak. The values i.e.\ for $U_{12}=0.5$ are
$\Delta\epsilon_{4}^{(0)}=0$ and $\Delta\epsilon_{5}^{(0)}=-0.5$, for
$U_{12}=0$ one obtains $\Delta\epsilon_{4}^{(0)}=-0.5$ and
$\Delta\epsilon_{5}^{(0)}=-1$.

\begin{figure*}
  \input{figures/fig7.tex}
  \caption{Like  Fig. \ref{fig:6}, but for $\e_1=\e_2=-1$, i.e.\ 
    the two local levels $\e_1$ and $\e_2$  are equally occupied
    and contribute symmetrically, i.e.\ to separate Kondo effects 
    at $U_{12}=0$.
  }
  \label{fig:7}
\end{figure*}
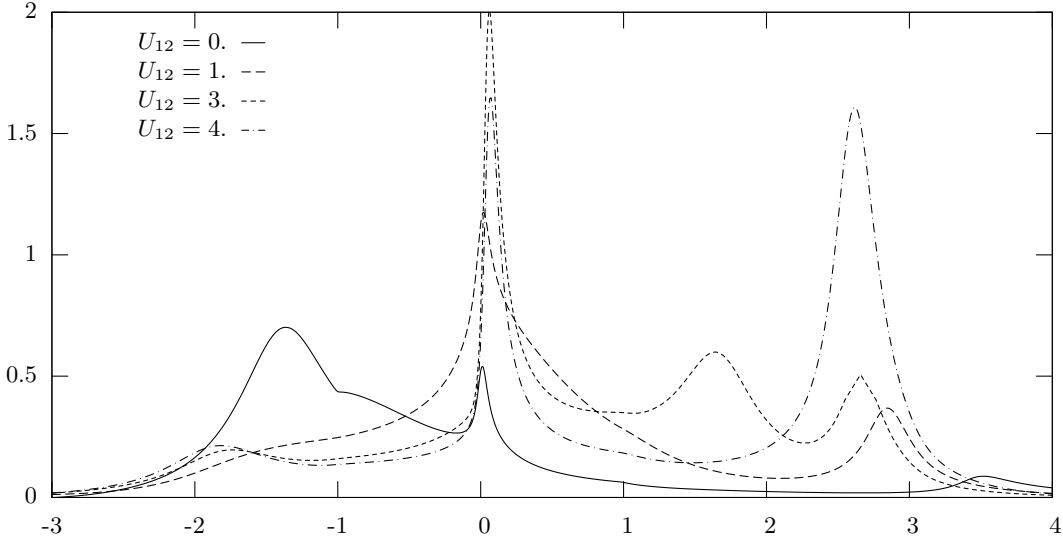
These considerations already give a rough explanation for the observed
shift of spectral weight, if one takes into account that peak
positions are usually renormalized away from the chemical potential by
interactions. The crossover region itself bears interesting physics in
it. At the point $U_{12}=0.5$, $\epsilon_{1}=-1$ and
$\epsilon_{2}=-0.5$ as before, the two states $E_{1;1}=\epsilon_{1}=-1$
and $E_{2;12}=\epsilon_1+\epsilon_2+U_{12}=-1$ with single and double
occupancy are degenerate ground states in zeroth order, which puts the
system in some kind of intermediate valence regime. This corresponds
well with a broad distribution of spectral weight near $\omega=0$. The effect
is seen in the curve for $U_{12}=1$, and even more pronounced, in the
following  Fig. \ref{fig:7} for $U_{12}=1$, with
$\epsilon_1=\epsilon_2=-1$. Consistently, in Fig. \ref{fig:7}, 
for vanishing
$U_{12}=0$ a marked Kondo effect with a thin ASR near $\omega=0$
reappears. In this case the ground state $E_{2;12}=-2$ is well separated
from the first excited states $E_{1;1}=E_{2;2}=\epsilon_1=-1$.

With respect to the local many-body correlations the crossover
scenario implies the following consequences: $U_{12}$ can be viewed as
a measure for the competition between tendencies to occupy each of the
two local levels. For large $U_{12}>U_{11}$ this competition favors
the lower local level $\epsilon_1$ and its corresponding
Kondo effect. Level $\epsilon_2$ then is only populated with a small
probability and furnishes only small side peaks near the
ASR. Correlations due to a virtual Kondo effect of level $\epsilon_2$
stay incomplete. With $U_{12}\approx U_{11}=U_{22}$ reaching the value
of the other Coulomb-matrix elements all of the different forms of
local double occupancy become energetically comparable, so that the
side peaks are strengthened somewhat due to increased interference
of the correlations. Below the crossover to the doubly occupied
ground state, however, individual Kondo effects for level $\epsilon_1$
and $\epsilon_2$ start to emerge. They do not show up for small
absolute values of $\epsilon_1$ and $\epsilon_2$, where ground states
with different local occupancy have comparable energies (on the scale
of the Anderson width) and low-lying charge fluctuations cause a broad
peak near $\omega=0$. If $\epsilon_1$ and $\epsilon_2$ are both well
below the chemical potential a complete decoupling then occurs at the
point $U_{12}=0$, where each of the levels only interacts with its own
band of the same symmetry. Due to different level positions the two
Kondo temperatures $T_{Kj}\sim \exp(-\frac{1}{J_j \rho_{\sigma}^{(0)}(0)})$,
$J_j \approx \frac{V^2}{|\epsilon_j|}+\frac{V^2}{\epsilon_j+U_{jj}}$ 
may come out largely different,
as a possibility. With e.g.\ $T_{K_1}<T<T_{K_2}$  the ASR due to $\epsilon_2$
is completely formed, whereas the one due to $\epsilon_1$ is still in an
early stage; the low-energy excitations then would be largely
determined by the Kondo effect of the higher lying local level. 

For
our set of parameters $T_{K_2}$ should be considerably higher than
$T_{K_1}$, and the position of $\epsilon_2=-0.5$ indicates an
intermediate valence region. The corresponding ``ASR'' then
significantly broadens and acquires some degree of linewidth via
charge fluctuations and simple one-particle scattering $\propto
V^2$. Thus due to the role of temperature variations relative to
$T_{K_1}$ and  $T_{K_2}$ and due to different 
level positions $\epsilon_2$ in the interval between $\epsilon_1$ and
the chemical potential $\mu=0$, the crossover scenario presented here
can involve quite different shapes of the one-particle excitation
spectrum. In particular, when even more complicated multi-orbital
models are used e.g.\ in calculations of correlated bandstructures,
great care seems necessary to understand and control all parts of the
calculation and the physical relevance of all features. We will
elaborate on this cautionary remark below also in connection with some
technical aspects of the approximations used.

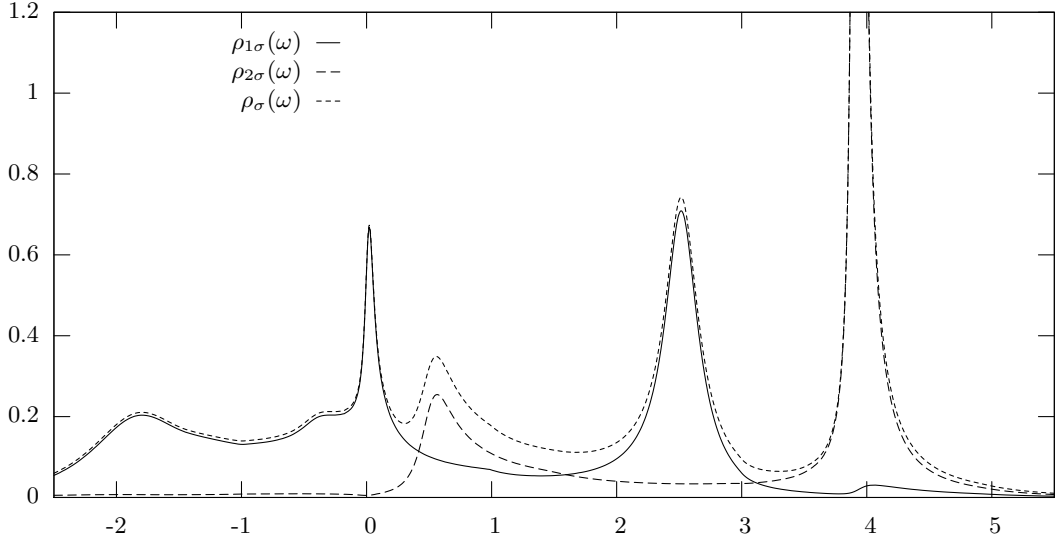
\begin{figure*}
  \input{figures/fig8.tex}
  \caption{Partial local one-particle excitation spectra $\rho_{1
      \sigma}(\omega)$ ,
    $\rho_{2 \sigma}(\omega)$ in comparison with the combined spectrum $\rho_{\sigma}(\omega)=\rho_{1
      \sigma}(\omega)+\rho_{2  \sigma}(\omega)$. The model parameters are as in
    Fig. \ref{fig:6} with $U_{12}=4$.}
  \label{fig:8}
\end{figure*}
With  Fig. \ref{fig:8} we present further evidence to support our picture of the
crossover scenario. We have resolved spectra from  Fig. \ref{fig:6} on each site
of the crossover, into the partial spectra, which incorporate only
the excitation on one of the local levels.  Fig. \ref{fig:8} clearly demonstrates
the strong dominance of the Kondo effect on the lower level
$\epsilon_1$ when $U_{12}$ is large: The ASR seen for $U_{12}=4$ is
completely contained in the partial spectrum $\rho_{1 \sigma}(\omega)$,
whereas $\rho_{2 \sigma}(\omega)$ is continuously flat near $\omega=0$,
although $T=\frac{1}{50 k_B}$ is of the order of a hypothetical
$T_{K_2}$ of an isolated level $\epsilon_2$. In  Fig. \ref{fig:7} on the other
hand, 
we have seen a symmetrical situation with
$\epsilon_1=\epsilon_2=-1$. The ASR, which shows up for $U_{12}=0$,
is clearly shared in equal parts by both partial spectra. A similar effect
in the spectrum with lowest $U_{12}$ of  Fig. \ref{fig:6} is not seen
due to
the different nature of correlations in the Intermediate Valence
regime. 

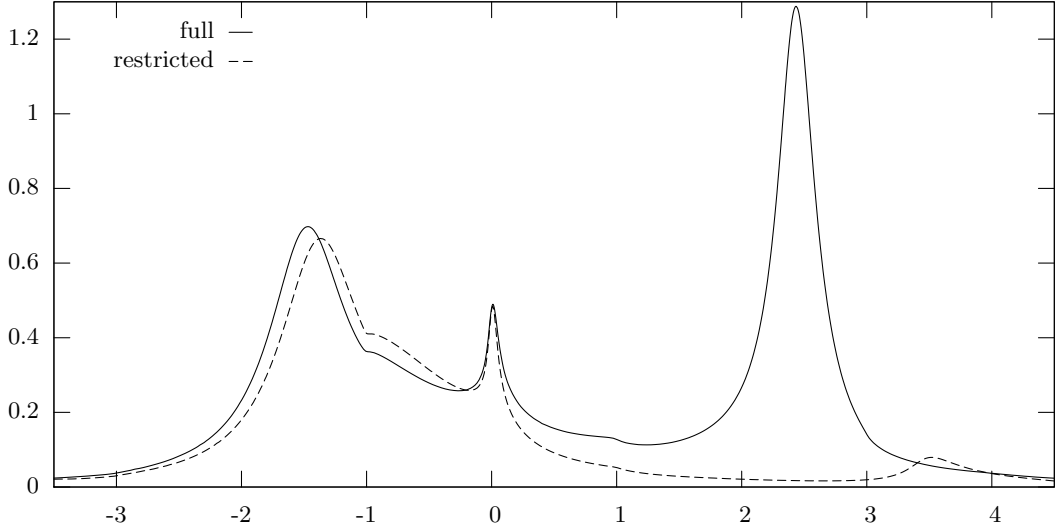
\begin{figure*}
  \input{figures/fig9.tex}
  \caption{Local one-particle excitation with restricted (local
    occupancy $\leq 2$) and full (local occupancy $\leq 4$).
    The model parameters are as in Fig. \ref{fig:7} with  $U_{12}=0$.}
  \label{fig:9}
\end{figure*}

The regime with the doubly occupied local ground state is apparently
not as well described by our ENCACF-approximation (not to mention the
simpler SNCACF) as the former regime with the singly occupied local
ground state. A plain hint  comes from the lack of spectral
intensity in the high energy region above $\omega=0$. More subtle is a
corresponding influence on the correlations determining the spectrum
nearby the chemical potential. With a local ground state derived from
$E_{2;12}=\epsilon_1+\epsilon_2+U_{12}$ and $U_{12}$ small, excitations
to a local state with triple occupancy and energy $\epsilon_1+2
\epsilon_2+U_{12}+U_{22}$, i.e.\ $\Delta
\epsilon_{7}^{(0)}=\epsilon_2+U_{22}$, or energy $2 \epsilon_1+
\epsilon_2+U_{12}+U_{11}$, i.e.\ $\Delta
\epsilon_{8}^{(0)}=\epsilon_1+U_{11}$, contribute to the
dynamics. Their importance compares with that of $\Delta
\epsilon_{4}^{(0)}=\epsilon_2+U_{12}$ and $\Delta
\epsilon_{5}^{(0)}=\epsilon_1+U_{12}$, similar to the importance of
both one-particle excitations at $\epsilon_1$ and $\epsilon_1+U$ for
an Anderson model with finite $U$. This deficiency can only be cured,
when the Hilbert space underlying the calculations is enlarged to
account for triple ( and quadruple)  local occupancy. Thereby the
system of integral equations for propagators again becomes
considerably enlarged. 

We have also solved the correspondingly
completed ENCACF-theory numerically and present a spectrum in  
Fig. \ref{fig:9}.
It is calculated for the most interesting case of
$\epsilon_1=\epsilon_2=-1$, $U_{12}=0$ and $U_{11}=U_{22}=3$,
$\beta=50$ and is compared with the incomplete spectrum discussed
before. The difference, particularly in the high-energy region is
clearly visible. On the other hand these differences for cases, where the local
ground state is singly occupied, essentially only show up at very high
excitations energies $\omega>2 U$ and can safely be ignored.

A second technical remark also refers to our note of caution made
above. Although from a   puristic  point of view and in agreement with
physical insight, the system for $U_{12}=0$ decouples dynamically into
two independent ones, this is in general not realized in approximate 
version of  direct perturbation theory,  which from the beginning incorporates
$U_{12}$
into the  unperturbed local system. For example, with
$U_{12}=0$ and $\epsilon_j< -\Delta_j < 0 < \Delta_j < \epsilon_j
+U_{jj}$ $(j=1,2)$, one should observe two independent
Kondo effects. Although as we have demonstrated, this is actually
found for large enough absolute values of both $\epsilon_j$, we are
not so confident about our findings in the crossover regime. The
point can be visualized with  Fig. \ref{fig:1}(a). The third self-energy diagram
shown there has an intermediate state $n_1=n_2=1$, which is the
ground state for small $U_{12}$. Fluctuations like the one shown via
the downrunning band electron line, as projected on the dynamics of
level $\epsilon_1$, can only take place when an electron is present in
level $\epsilon_2$. This establishes a spurious kind of correlation,
which only is eliminated from the calculation when all of the possible
vertex corrections are actually taken into account in the vertex shown
at the top. Only this can lead to all combinations of individual time
orders, which together would decouple the two subsystems. 


Our ENCACF
captures the leading part thereof via the vertex corrections shown in
figure~\ref{fig:1}(b). It is, however, not settled yet how important higher
contributions with more crossings are in the different regimes. Although
there is some experience for the simple Anderson model \cite{anders:PostNCA95,kroha:CTMA97,greweCA108},
suggesting  a qualitatively well behaved ENCA, in particular the crossover
regime found here might need even better approximations.


\section{Conclusion and Perspective}
\label{sec:4}
 
Our investigation of a multi-orbital model with a new crossing
approximation, the ENCACF, which goes beyond a simple modification of
NCA and works for all finite values of Coulomb parameters, has revealed
a variety of possible forms of one particle excitation spectra.  We
have studied the role of different model parameters, temperature $T$,
energies $\epsilon_j$ of the two local one-particle levels with
different symmetry, and of the Coulomb-matrix elements $U_{jl}$ between
electrons in level $\epsilon_j$ and $\epsilon_l$. Different regimes
could be characterized, according to the nature of the local
ground state and of the local correlations. These regimes exhibit
different modified forms of a Kondo effect, when the levels
$\epsilon_1$ and $\epsilon_2$ are below the chemical potential and
representative values of $U_{jl}$ are large enough.

Single particle
transition between local states on the one hand and traces of
many-body correlations on the other hand could both be identified in
the calculated spectra and were discussed in a consistent
framework. The crossover between regimes of different local
groundstate occupancy involves interesting Intermediate Valence
phenomena, where charge fluctuations intervene. We have pointed to the
intricate physics, which determines the crossover region, and to some
problems of approximations which possibly render a complete
understanding difficult with the methods developed so far. Particular
caution seems necessary when interpreting  e.g.\ experimental data
from photoelectron spectroscopy with insufficient approximations,
using e.g.\ simpler forms of direct perturbation theory like NCA,
combined with the limit of infinite local Coulomb repulsion. 
Whereas  we believe, that some correct qualitative information can be
gathered in this way and a correct overall picture be developed, our
new calculations show, based on an essentially improved theory, show that
quantitative fits may be very misleading: The relevant
many-body scale, for example grossly differs and important spectral
information from the high energy region gets lost. Moreover, even the
detailed spectral intensities in the low energy region depend on the
rest of the spectrum and may come out incorrectly, in particular in
real physical parameter regimes of transition metal ions which do not
allow for an argument of universality. Nevertheless, our results also
support some earlier findings, in particular for the spectroscopy of
metallic compounds with Ce-atoms. 
For the use of multi-orbital models
of the kind studied in connection with self-consistent methods 
for correlated band structures, on the other hand, our
investigation furnishes a rather poor perspective due to the
complexity of the problem and the insufficient quality of suitable impurity
solvers proposed so far. Here we hope that generalizations of our
ENCACF may be helpful.


In any case, multi-orbital impurity solvers for lattice systems with
strong local correlations in most cases should be planned in an even
more general form than our model of equation~(\ref{hamilton}). 
The reason simply is that
symmetry decompositions of band states around different sites as
centers are not compatible : They combine the Bloch functions in a
different way. As a consequence, the medium around a given
representative site will mix the symmetry-decompositions,
i.e.\ $\epsilon_{1\underline{k}}$ and $\epsilon_{2\underline{k}}$ in our
model. In that case the possible diagrammatic processes make up a
larger set as before, with the consequence that the irreducible
self-energies of ionic propagators become a matrix. This matrix
involves also elements which are non-diagonal in the level indices
$j$. The same complication arises in e.g.\ the case, where an impurity
complex made up of two sites is inserted into a perfect crystal with
orbitals on each of the two sites. This again spoils the
conservation of a $j$ quantum number of local electrons and leads to
the same consequences as before. We have taken up studies of such
generalized multi-orbital models and will report on results in a
forthcoming paper.


\bibliographystyle{epj}
\bibliography{cef_paper.bib}

\end{document}

%% file: sig0_nca.tex
$\Sigma_{1\sigma}(z)=$
\parbox{28mm}{
    \begin{fmfgraph*}(14,30)
      \fmfv{decor.shape=triangle,decor.angle=30,decor.fill=0}{vo}
      \fmfdot{vu}
      \fmfbottom{u} \fmftop{o}
      \fmf{wiggly_arrow,tension=3,label=$1\sigma$,label.side=left}{u,vu}
      \fmflabel{$z$}{o}
      \fmf{dashes,label=$0$, label.side=left}{vu,vo}
      \fmf{wiggly_arrow,tension=3,label=$1\sigma$,label.side=left}{vo,o}
      \fmf{fermion,right=1,tension=0.2, label=$1\underline{k}\sigma$,label.side=right}{vu,vo}
      \fmffreeze
   \end{fmfgraph*}
 }
 $+\quad $
\parbox{28mm}{
    \begin{fmfgraph*}(14,30)
      \fmfv{decor.shape=triangle,decor.angle=30,decor.fill=1}{vo}
      \fmfdot{vu}
      \fmfbottom{u} \fmftop{o}
      \fmf{wiggly_arrow,tension=3,label=$1\sigma$,label.side=left}{u,vu}
      \fmflabel{$z$}{o}
      \fmf{dbl_wiggly_arrow,label=$\begin{matrix} 1\sigma\cr 1-\!\sigma\end{matrix}$, label.side=left}{vu,vo}
      \fmf{wiggly_arrow,tension=3,label=$1\sigma$,label.side=left}{vo,o}
      \fmf{fermion,left=1,tension=0.2, label=$1\underline{k}-\!\sigma$,label.side=left}{vo,vu}
      \fmffreeze
   \end{fmfgraph*}
 }
$ +\quad $
\parbox{25mm}{
    \begin{fmfgraph*}(14,30)
      \fmfv{decor.shape=triangle,decor.angle=30,decor.fill=0.5}{vo}
      \fmfdot{vu}
      \fmfbottom{u} \fmftop{o}
      \fmf{wiggly_arrow,tension=3,label=$1\sigma$,label.side=left}{u,vu}
      \fmflabel{$z$}{o}
      \fmf{dbl_wiggly_arrow,label=$\begin{matrix} 1\sigma\cr 2\sigma'\end{matrix}$, label.side=left}{vu,vo}
      \fmf{wiggly_arrow,tension=3,label=$1\sigma$,label.side=left}{vo,o}
      \fmf{fermion,left=1,tension=0.2, label=$2\underline{k}\sigma'$,label.side=left}{vo,vu}
      \fmffreeze
   \end{fmfgraph*}
 }%

%% file: vertex_enca.tex
\parbox{20mm}{
  \begin{fmfgraph*}(12,25)
    \fmfv{decor.shape=triangle,decor.angle=30,decor.fill=0}{v1}
    \fmfbottom{u} \fmftop{o}
    \fmf{dashes,label=$0$,label.side=left}{u,v1}
    \fmf{wiggly_arrow,label=$1\sigma$,label.side=left}{v1,o}
    \fmflabel{$z$}{o}
    \fmffreeze
    \fmfright{r}
    \fmfforce{(0.5*w+220,0.5*h)}{r}
    \fmf{plain_arrow,label=$1\underline{k}\sigma$,label.side=right}{r,v1}
  \end{fmfgraph*}
}
=
\parbox{15mm}{
  \begin{fmfgraph}(12,25)
    \fmfdot{v1}
    \fmfbottom{u} \fmftop{o}
    \fmf{dashes}{u,v1}
    \fmf{wiggly_arrow}{v1,o}
    \fmflabel{$z$}{o}
    \fmffreeze
    \fmfright{r}
    \fmfforce{(0.5*w+180,0.5*h)}{r}
    \fmf{plain_arrow}{r,v1}
  \end{fmfgraph}
}
$\quad +\quad$
\parbox{28mm}{
  \begin{fmfgraph*}(15,40)
    \fmfdot{vu,vm,vo}
    \fmfbottom{u} \fmftop{o}
    \fmf{dashes,tension=2}{u,vu}
    \fmf{wiggly_arrow,label=$1-\!\sigma$,label.side=left}{vu,vm}
    \fmf{dbl_wiggly_arrow,label=$\begin{matrix} 1\sigma\cr 1-\!\sigma\end{matrix}$,label.side=left}{vm,vo}
    \fmf{fermion,left=0.8,tension=0.2}{vo,vu}
    \fmf{wiggly_arrow,tension=2}{vo,o}
    \fmflabel{$z$}{o}
    \fmffreeze
    \fmfright{r}
    \fmfforce{(0.5*w+400,0.5*h)}{r}
    \fmf{plain}{vm,vr1}
    \fmf{plain}{vr1,vr2}
    \fmf{plain_arrow}{vr2,r}
    \fmffreeze
   \fmfiv{l=$1\underline{k}'-\!\sigma$,l.a=0,l.d=0.5w}{vloc(__vo)}
  \end{fmfgraph*}
}
$\quad +\quad$
\parbox{28mm}{
  \begin{fmfgraph*}(15,40)
    \fmfdot{vu,vm,vo}
    \fmfbottom{u} \fmftop{o}
    \fmf{dashes,tension=2}{u,vu}
    \fmf{wiggly_arrow,label=$2\sigma'$,label.side=left}{vu,vm}
    \fmf{dbl_wiggly_arrow,label=$\begin{matrix} 1\sigma\cr 2\sigma'\end{matrix}$,label.side=left}{vm,vo}
    \fmf{fermion,left=0.8,tension=0.2}{vo,vu}
    \fmf{wiggly_arrow,tension=2}{vo,o}
    \fmflabel{$z$}{o}
    \fmffreeze
    \fmfright{r}
    \fmfforce{(0.5*w+400,0.5*h)}{r}
    \fmf{plain}{vm,vr1}
    \fmf{plain}{vr1,vr2}
    \fmf{plain_arrow}{vr2,r}
    \fmffreeze
   \fmfiv{l=$2\underline{k}'\sigma'$,l.a=0,l.d=0.5w}{vloc(__vo)}
  \end{fmfgraph*}
}%

%% file: figures/fig2all.tex
\begin{picture}(0,0)%
\includegraphics{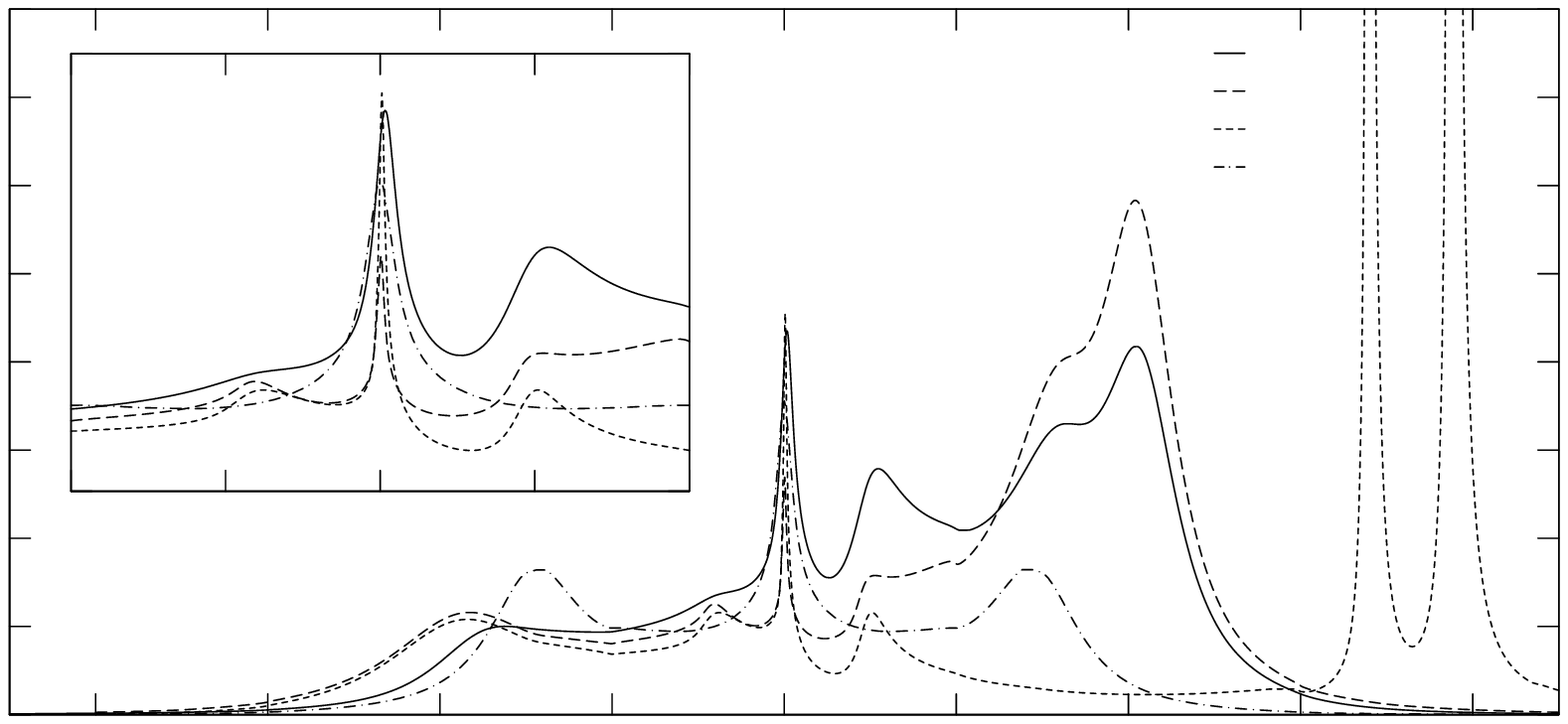}%
\end{picture}%
\begingroup
\setlength{\unitlength}{0.0200bp}%
\begin{picture}(25199,11880)(0,0)%
\put(1650,1100){\makebox(0,0)[r]{\strut{} 0}}%
\put(1650,2379){\makebox(0,0)[r]{\strut{} 0.2}}%
\put(1650,3658){\makebox(0,0)[r]{\strut{} 0.4}}%
\put(1650,4936){\makebox(0,0)[r]{\strut{} 0.6}}%
\put(1650,6215){\makebox(0,0)[r]{\strut{} 0.8}}%
\put(1650,7494){\makebox(0,0)[r]{\strut{} 1}}%
\put(1650,8772){\makebox(0,0)[r]{\strut{} 1.2}}%
\put(1650,10051){\makebox(0,0)[r]{\strut{} 1.4}}%
\put(1650,11330){\makebox(0,0)[r]{\strut{} 1.6}}%
\put(3172,550){\makebox(0,0){\strut{}-4}}%
\put(5667,550){\makebox(0,0){\strut{}-3}}%
\put(8161,550){\makebox(0,0){\strut{}-2}}%
\put(10656,550){\makebox(0,0){\strut{}-1}}%
\put(13150,550){\makebox(0,0){\strut{} 0}}%
\put(15644,550){\makebox(0,0){\strut{} 1}}%
\put(18139,550){\makebox(0,0){\strut{} 2}}%
\put(20633,550){\makebox(0,0){\strut{} 3}}%
\put(23128,550){\makebox(0,0){\strut{} 4}}%
\put(19111,10691){\makebox(0,0)[r]{\strut{}ENCACF $U=2.$, $\beta=50.$}}%
\put(19111,10141){\makebox(0,0)[r]{\strut{}SNCACF $U=2.$, $\beta=200.$}}%
\put(19111,9591){\makebox(0,0)[r]{\strut{}SNCACF $U=4.$, $\beta=200.$}}%
\put(19111,9041){\makebox(0,0)[r]{\strut{}ENCA $U=2.$, $\beta=50.$}}%
\put(2540,4340){\makebox(0,0)[r]{\strut{} 0}}%
\put(2540,10682){\makebox(0,0)[r]{\strut{} 1}}%
\put(2815,3790){\makebox(0,0){\strut{}-1}}%
\put(5055,3790){\makebox(0,0){\strut{}-0.5}}%
\put(7295,3790){\makebox(0,0){\strut{} 0}}%
\put(9535,3790){\makebox(0,0){\strut{} 0.5}}%
\put(11775,3790){\makebox(0,0){\strut{} 1}}%
\end{picture}%
\endgroup
 

%% file: figures/fig3a.tex
\begin{picture}(0,0)%
\includegraphics{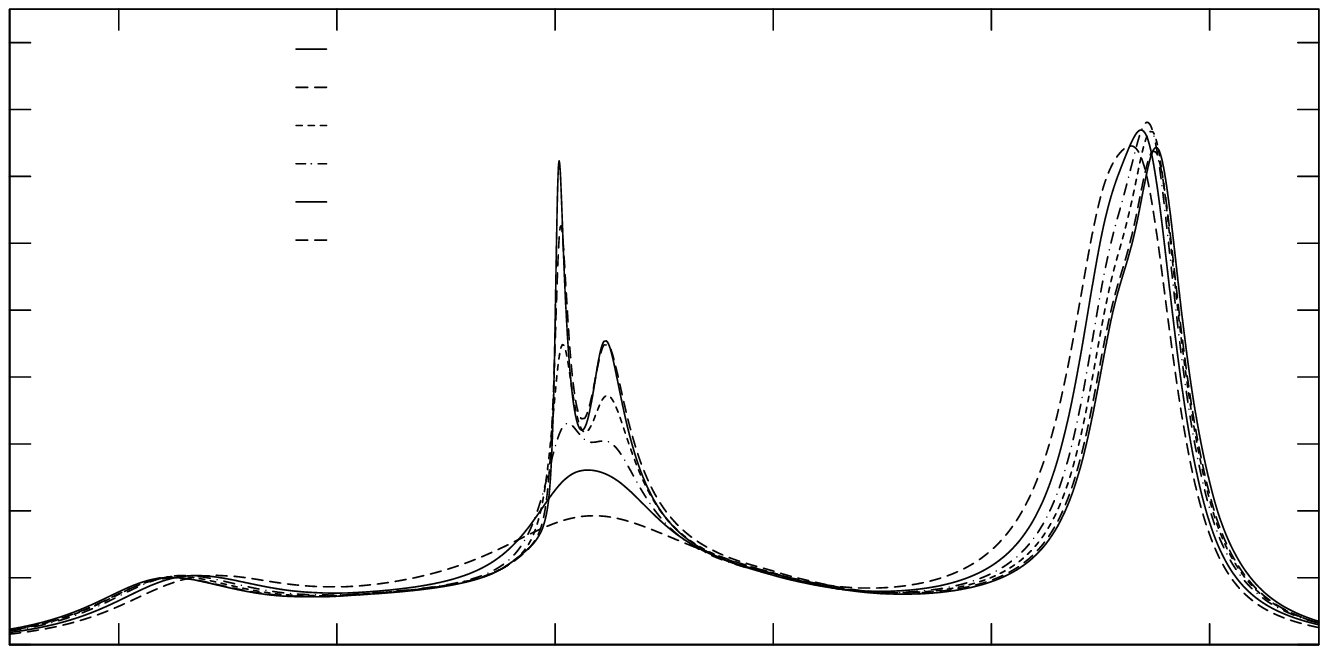}%
\end{picture}%
\begingroup
\setlength{\unitlength}{0.0200bp}%
\begin{picture}(21600,10800)(0,0)%
\put(1650,1100){\makebox(0,0)[r]{\strut{} 0}}%
\put(1650,2063){\makebox(0,0)[r]{\strut{} 0.2}}%
\put(1650,3026){\makebox(0,0)[r]{\strut{} 0.4}}%
\put(1650,3989){\makebox(0,0)[r]{\strut{} 0.6}}%
\put(1650,4953){\makebox(0,0)[r]{\strut{} 0.8}}%
\put(1650,5916){\makebox(0,0)[r]{\strut{} 1}}%
\put(1650,6879){\makebox(0,0)[r]{\strut{} 1.2}}%
\put(1650,7842){\makebox(0,0)[r]{\strut{} 1.4}}%
\put(1650,8805){\makebox(0,0)[r]{\strut{} 1.6}}%
\put(1650,9768){\makebox(0,0)[r]{\strut{} 1.8}}%
\put(3496,550){\makebox(0,0){\strut{}-2}}%
\put(6637,550){\makebox(0,0){\strut{}-1}}%
\put(9779,550){\makebox(0,0){\strut{} 0}}%
\put(12921,550){\makebox(0,0){\strut{} 1}}%
\put(16062,550){\makebox(0,0){\strut{} 2}}%
\put(19204,550){\makebox(0,0){\strut{} 3}}%
\put(5775,9675){\makebox(0,0)[r]{\strut{}$\beta=100.$}}%
\put(5775,9125){\makebox(0,0)[r]{\strut{}$\beta=50.$}}%
\put(5775,8575){\makebox(0,0)[r]{\strut{}$\beta=25.$}}%
\put(5775,8025){\makebox(0,0)[r]{\strut{}$\beta=12.5$}}%
\put(5775,7475){\makebox(0,0)[r]{\strut{}$\beta=6.25$}}%
\put(5775,6925){\makebox(0,0)[r]{\strut{}$\beta=3.125$}}%
\end{picture}%
\endgroup
 

%% file: figures/fig3b.tex
\begin{picture}(0,0)%
\includegraphics{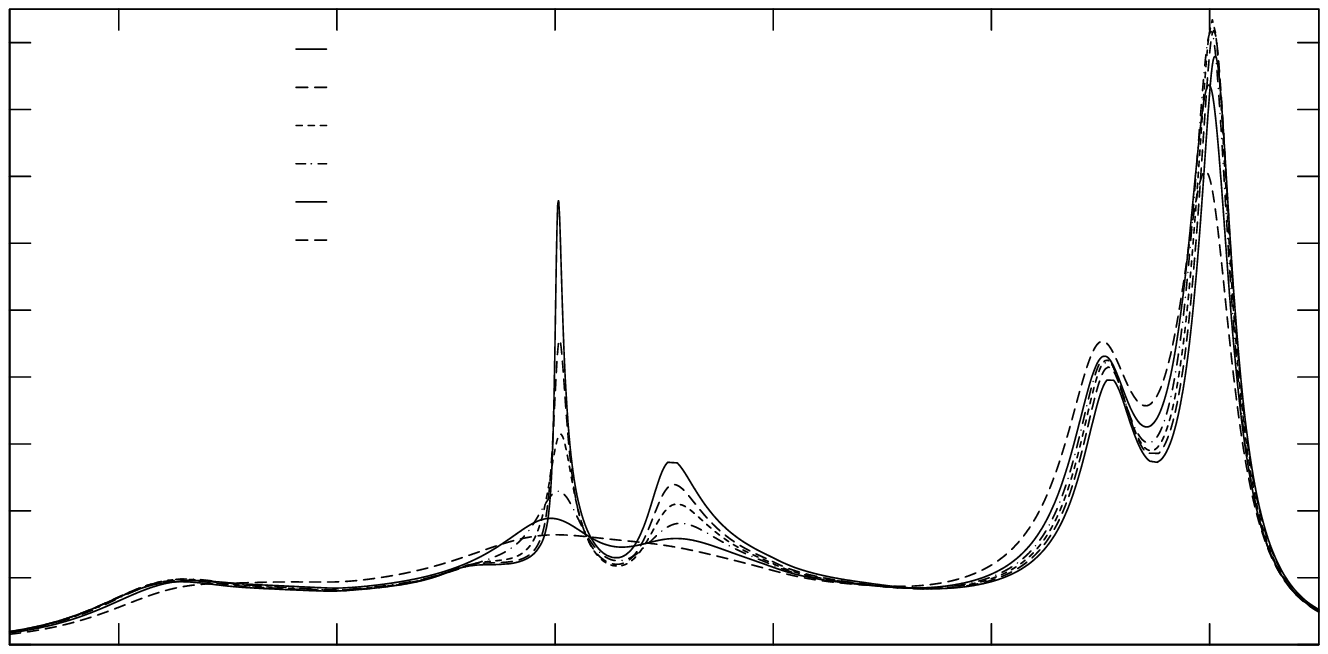}%
\end{picture}%
\begingroup
\setlength{\unitlength}{0.0200bp}%
\begin{picture}(21600,10800)(0,0)%
\put(1650,1100){\makebox(0,0)[r]{\strut{} 0}}%
\put(1650,2063){\makebox(0,0)[r]{\strut{} 0.2}}%
\put(1650,3026){\makebox(0,0)[r]{\strut{} 0.4}}%
\put(1650,3989){\makebox(0,0)[r]{\strut{} 0.6}}%
\put(1650,4953){\makebox(0,0)[r]{\strut{} 0.8}}%
\put(1650,5916){\makebox(0,0)[r]{\strut{} 1}}%
\put(1650,6879){\makebox(0,0)[r]{\strut{} 1.2}}%
\put(1650,7842){\makebox(0,0)[r]{\strut{} 1.4}}%
\put(1650,8805){\makebox(0,0)[r]{\strut{} 1.6}}%
\put(1650,9768){\makebox(0,0)[r]{\strut{} 1.8}}%
\put(3496,550){\makebox(0,0){\strut{}-2}}%
\put(6637,550){\makebox(0,0){\strut{}-1}}%
\put(9779,550){\makebox(0,0){\strut{} 0}}%
\put(12921,550){\makebox(0,0){\strut{} 1}}%
\put(16062,550){\makebox(0,0){\strut{} 2}}%
\put(19204,550){\makebox(0,0){\strut{} 3}}%
\put(5775,9675){\makebox(0,0)[r]{\strut{}$\beta=100.$}}%
\put(5775,9125){\makebox(0,0)[r]{\strut{}$\beta=50.$}}%
\put(5775,8575){\makebox(0,0)[r]{\strut{}$\beta=25.$}}%
\put(5775,8025){\makebox(0,0)[r]{\strut{}$\beta=12.5$}}%
\put(5775,7475){\makebox(0,0)[r]{\strut{}$\beta=6.25$}}%
\put(5775,6925){\makebox(0,0)[r]{\strut{}$\beta=3.125$}}%
\end{picture}%
\endgroup
 

%% file: figures/fig4.tex
\begin{picture}(0,0)%
\includegraphics{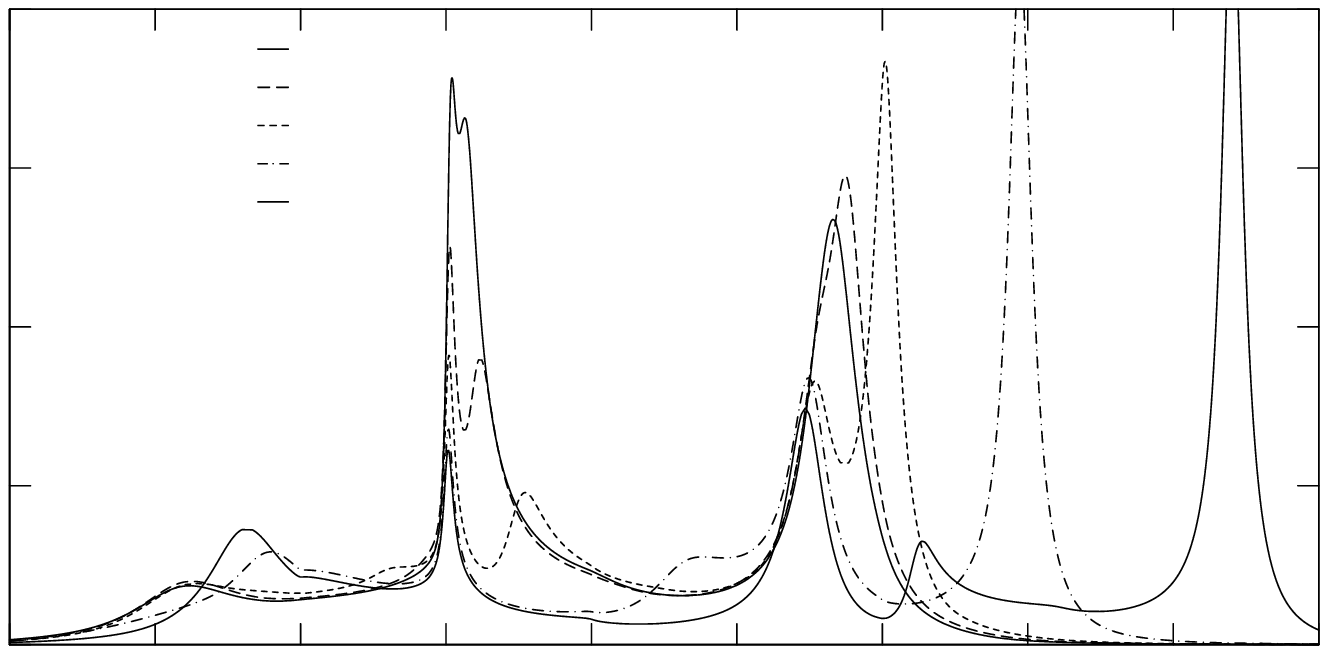}%
\end{picture}%
\begingroup
\setlength{\unitlength}{0.0200bp}%
\begin{picture}(21600,10800)(0,0)%
\put(1650,1100){\makebox(0,0)[r]{\strut{} 0}}%
\put(1650,3388){\makebox(0,0)[r]{\strut{} 0.5}}%
\put(1650,5675){\makebox(0,0)[r]{\strut{} 1}}%
\put(1650,7963){\makebox(0,0)[r]{\strut{} 1.5}}%
\put(1650,10250){\makebox(0,0)[r]{\strut{} 2}}%
\put(1925,550){\makebox(0,0){\strut{}-3}}%
\put(4019,550){\makebox(0,0){\strut{}-2}}%
\put(6114,550){\makebox(0,0){\strut{}-1}}%
\put(8208,550){\makebox(0,0){\strut{} 0}}%
\put(10303,550){\makebox(0,0){\strut{} 1}}%
\put(12397,550){\makebox(0,0){\strut{} 2}}%
\put(14492,550){\makebox(0,0){\strut{} 3}}%
\put(16586,550){\makebox(0,0){\strut{} 4}}%
\put(18681,550){\makebox(0,0){\strut{} 5}}%
\put(20775,550){\makebox(0,0){\strut{} 6}}%
\put(5225,9675){\makebox(0,0)[r]{\strut{}$\e_2=-0.9$}}%
\put(5225,9125){\makebox(0,0)[r]{\strut{}$\e_2=-0.8$}}%
\put(5225,8575){\makebox(0,0)[r]{\strut{}$\e_2=-0.5$}}%
\put(5225,8025){\makebox(0,0)[r]{\strut{}$\e_2=+0.5$}}%
\put(5225,7475){\makebox(0,0)[r]{\strut{}$\e_2=+2.0$}}%
\end{picture}%
\endgroup
 

%% file: figures/fig5.tex
\begin{picture}(0,0)%
\includegraphics{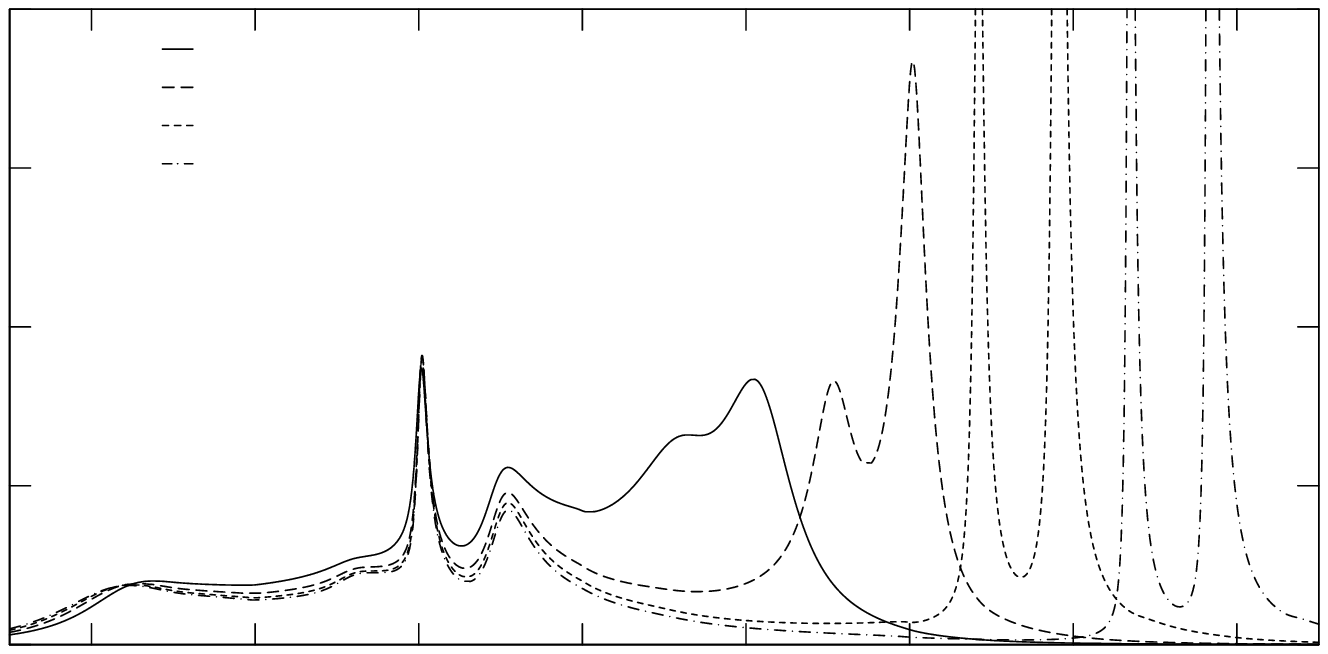}%
\end{picture}%
\begingroup
\setlength{\unitlength}{0.0200bp}%
\begin{picture}(21600,10800)(0,0)%
\put(1650,1100){\makebox(0,0)[r]{\strut{} 0}}%
\put(1650,3388){\makebox(0,0)[r]{\strut{} 0.5}}%
\put(1650,5675){\makebox(0,0)[r]{\strut{} 1}}%
\put(1650,7963){\makebox(0,0)[r]{\strut{} 1.5}}%
\put(1650,10250){\makebox(0,0)[r]{\strut{} 2}}%
\put(3103,550){\makebox(0,0){\strut{}-2}}%
\put(5459,550){\makebox(0,0){\strut{}-1}}%
\put(7816,550){\makebox(0,0){\strut{} 0}}%
\put(10172,550){\makebox(0,0){\strut{} 1}}%
\put(12528,550){\makebox(0,0){\strut{} 2}}%
\put(14884,550){\makebox(0,0){\strut{} 3}}%
\put(17241,550){\makebox(0,0){\strut{} 4}}%
\put(19597,550){\makebox(0,0){\strut{} 5}}%
\put(3850,9675){\makebox(0,0)[r]{\strut{}$U=2.$}}%
\put(3850,9125){\makebox(0,0)[r]{\strut{}$U=3.$}}%
\put(3850,8575){\makebox(0,0)[r]{\strut{}$U=4.$}}%
\put(3850,8025){\makebox(0,0)[r]{\strut{}$U=5.$}}%
\end{picture}%
\endgroup
 

%% file: figures/fig6.tex
\begin{picture}(0,0)%
\includegraphics{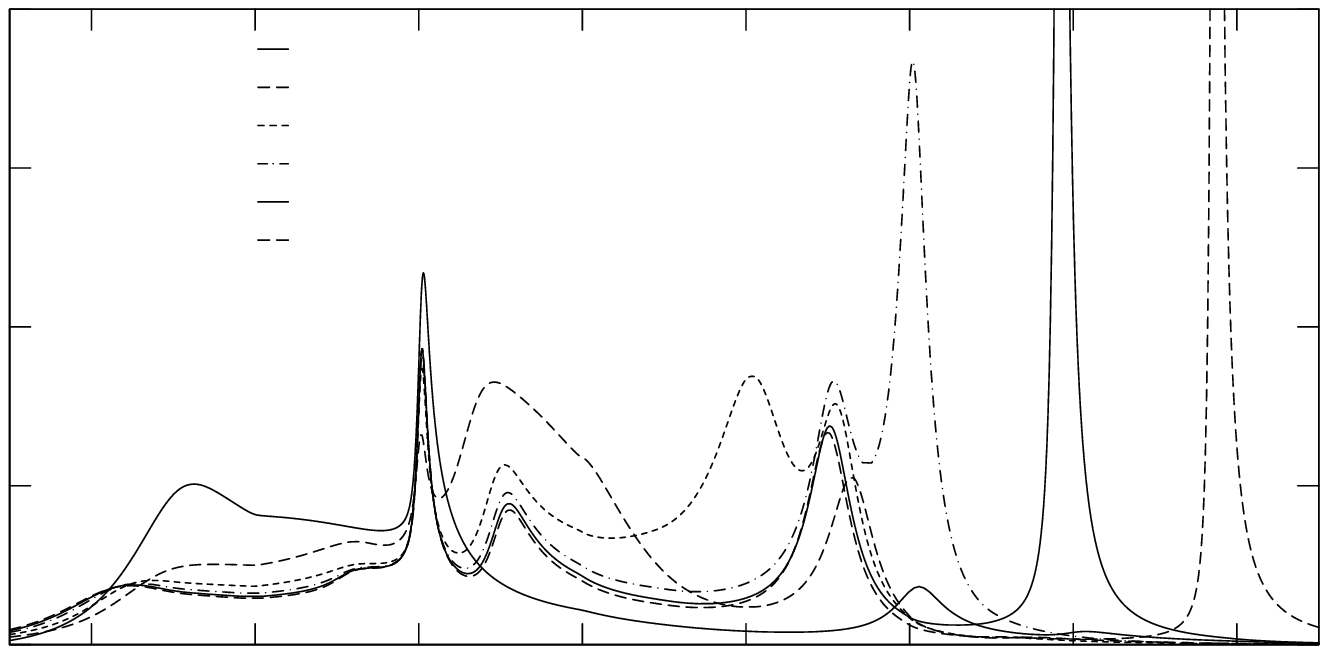}%
\end{picture}%
\begingroup
\setlength{\unitlength}{0.0200bp}%
\begin{picture}(21600,10800)(0,0)%
\put(1650,1100){\makebox(0,0)[r]{\strut{} 0}}%
\put(1650,3388){\makebox(0,0)[r]{\strut{} 0.5}}%
\put(1650,5675){\makebox(0,0)[r]{\strut{} 1}}%
\put(1650,7963){\makebox(0,0)[r]{\strut{} 1.5}}%
\put(1650,10250){\makebox(0,0)[r]{\strut{} 2}}%
\put(3103,550){\makebox(0,0){\strut{}-2}}%
\put(5459,550){\makebox(0,0){\strut{}-1}}%
\put(7816,550){\makebox(0,0){\strut{} 0}}%
\put(10172,550){\makebox(0,0){\strut{} 1}}%
\put(12528,550){\makebox(0,0){\strut{} 2}}%
\put(14884,550){\makebox(0,0){\strut{} 3}}%
\put(17241,550){\makebox(0,0){\strut{} 4}}%
\put(19597,550){\makebox(0,0){\strut{} 5}}%
\put(5225,9675){\makebox(0,0)[r]{\strut{}$U_{12}=0.$}}%
\put(5225,9125){\makebox(0,0)[r]{\strut{}$U_{12}=1.$}}%
\put(5225,8575){\makebox(0,0)[r]{\strut{}$U_{12}=2.$}}%
\put(5225,8025){\makebox(0,0)[r]{\strut{}$U_{12}=3.$}}%
\put(5225,7475){\makebox(0,0)[r]{\strut{}$U_{12}=4.$}}%
\put(5225,6925){\makebox(0,0)[r]{\strut{}$U_{12}=5.$}}%
\end{picture}%
\endgroup
 

%% file: figures/fig7.tex
\begin{picture}(0,0)%
\includegraphics{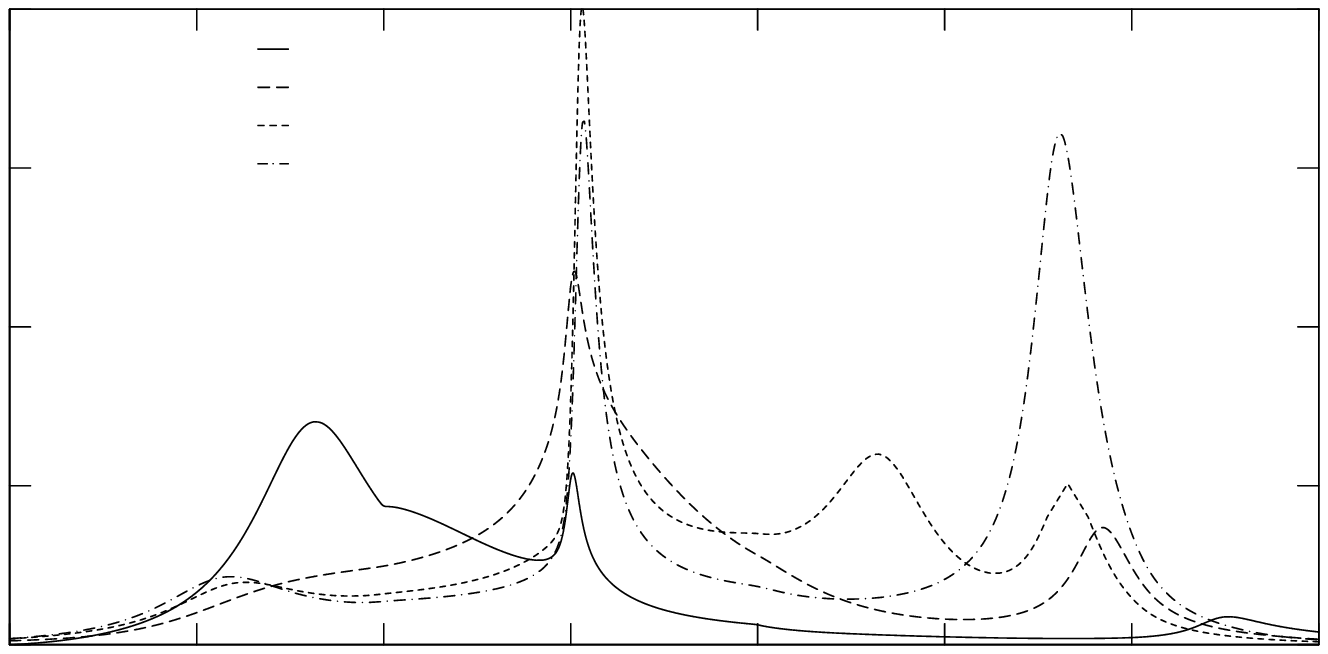}%
\end{picture}%
\begingroup
\setlength{\unitlength}{0.0200bp}%
\begin{picture}(21600,10800)(0,0)%
\put(1650,1100){\makebox(0,0)[r]{\strut{} 0}}%
\put(1650,3388){\makebox(0,0)[r]{\strut{} 0.5}}%
\put(1650,5675){\makebox(0,0)[r]{\strut{} 1}}%
\put(1650,7963){\makebox(0,0)[r]{\strut{} 1.5}}%
\put(1650,10250){\makebox(0,0)[r]{\strut{} 2}}%
\put(1925,550){\makebox(0,0){\strut{}-3}}%
\put(4618,550){\makebox(0,0){\strut{}-2}}%
\put(7311,550){\makebox(0,0){\strut{}-1}}%
\put(10004,550){\makebox(0,0){\strut{} 0}}%
\put(12696,550){\makebox(0,0){\strut{} 1}}%
\put(15389,550){\makebox(0,0){\strut{} 2}}%
\put(18082,550){\makebox(0,0){\strut{} 3}}%
\put(20775,550){\makebox(0,0){\strut{} 4}}%
\put(5225,9675){\makebox(0,0)[r]{\strut{}$U_{12}=0.$}}%
\put(5225,9125){\makebox(0,0)[r]{\strut{}$U_{12}=1.$}}%
\put(5225,8575){\makebox(0,0)[r]{\strut{}$U_{12}=3.$}}%
\put(5225,8025){\makebox(0,0)[r]{\strut{}$U_{12}=4.$}}%
\end{picture}%
\endgroup
 

%% file: figures/fig8.tex
\begin{picture}(0,0)%
\includegraphics{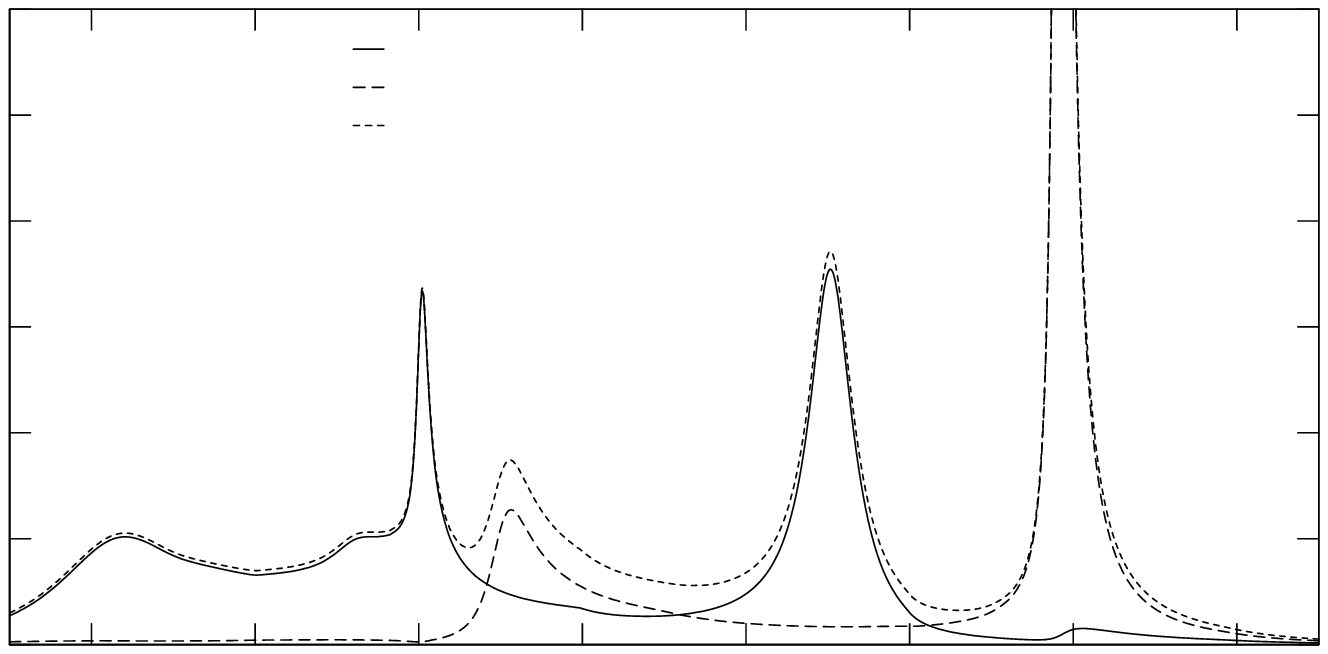}%
\end{picture}%
\begingroup
\setlength{\unitlength}{0.0200bp}%
\begin{picture}(21600,10800)(0,0)%
\put(1650,1100){\makebox(0,0)[r]{\strut{} 0}}%
\put(1650,2625){\makebox(0,0)[r]{\strut{} 0.2}}%
\put(1650,4150){\makebox(0,0)[r]{\strut{} 0.4}}%
\put(1650,5675){\makebox(0,0)[r]{\strut{} 0.6}}%
\put(1650,7200){\makebox(0,0)[r]{\strut{} 0.8}}%
\put(1650,8725){\makebox(0,0)[r]{\strut{} 1}}%
\put(1650,10250){\makebox(0,0)[r]{\strut{} 1.2}}%
\put(3103,550){\makebox(0,0){\strut{}-2}}%
\put(5459,550){\makebox(0,0){\strut{}-1}}%
\put(7816,550){\makebox(0,0){\strut{} 0}}%
\put(10172,550){\makebox(0,0){\strut{} 1}}%
\put(12528,550){\makebox(0,0){\strut{} 2}}%
\put(14884,550){\makebox(0,0){\strut{} 3}}%
\put(17241,550){\makebox(0,0){\strut{} 4}}%
\put(19597,550){\makebox(0,0){\strut{} 5}}%
\put(6600,9675){\makebox(0,0)[r]{\strut{}$\rho_{1\s}(\o)$}}%
\put(6600,9125){\makebox(0,0)[r]{\strut{}$\rho_{2\s}(\o)$}}%
\put(6600,8575){\makebox(0,0)[r]{\strut{}$\rho_{\s}(\o)$}}%
\end{picture}%
\endgroup
 

%% file: figures/fig9.tex
\begin{picture}(0,0)%
\includegraphics{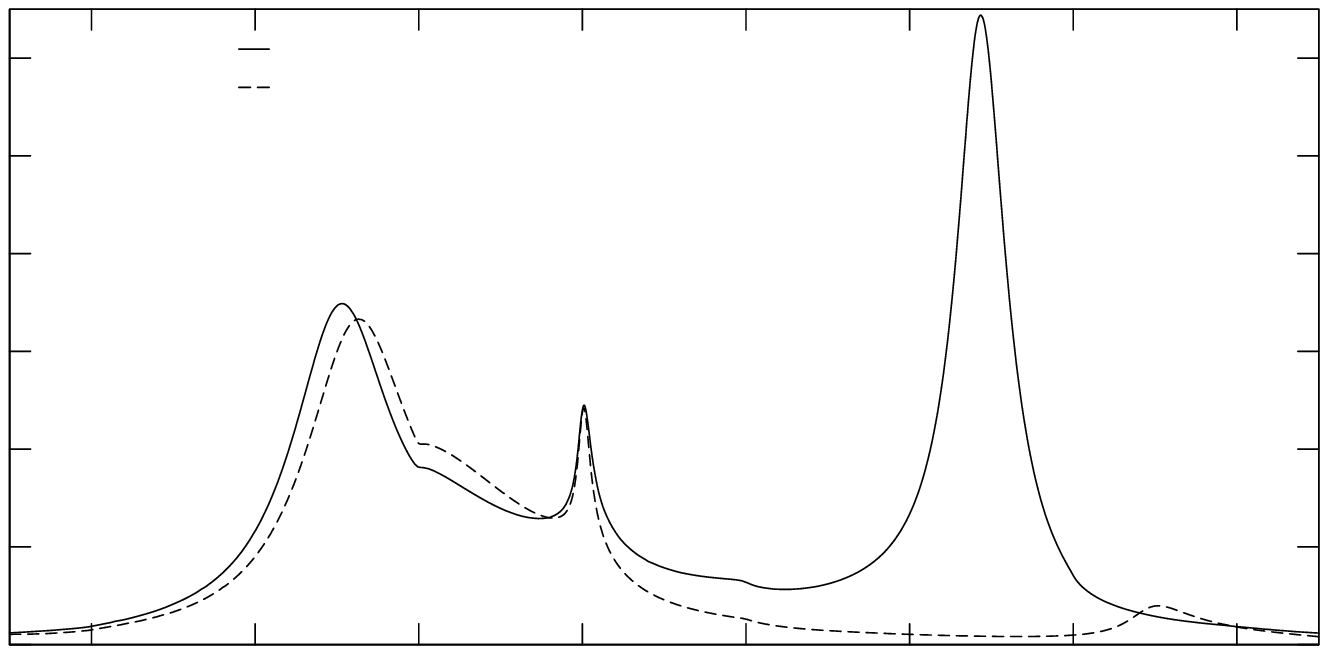}%
\end{picture}%
\begingroup
\setlength{\unitlength}{0.0200bp}%
\begin{picture}(21600,10800)(0,0)%
\put(1650,1100){\makebox(0,0)[r]{\strut{} 0}}%
\put(1650,2508){\makebox(0,0)[r]{\strut{} 0.2}}%
\put(1650,3915){\makebox(0,0)[r]{\strut{} 0.4}}%
\put(1650,5323){\makebox(0,0)[r]{\strut{} 0.6}}%
\put(1650,6731){\makebox(0,0)[r]{\strut{} 0.8}}%
\put(1650,8138){\makebox(0,0)[r]{\strut{} 1}}%
\put(1650,9546){\makebox(0,0)[r]{\strut{} 1.2}}%
\put(3103,550){\makebox(0,0){\strut{}-3}}%
\put(5459,550){\makebox(0,0){\strut{}-2}}%
\put(7816,550){\makebox(0,0){\strut{}-1}}%
\put(10172,550){\makebox(0,0){\strut{} 0}}%
\put(12528,550){\makebox(0,0){\strut{} 1}}%
\put(14884,550){\makebox(0,0){\strut{} 2}}%
\put(17241,550){\makebox(0,0){\strut{} 3}}%
\put(19597,550){\makebox(0,0){\strut{} 4}}%
\put(4950,9675){\makebox(0,0)[r]{\strut{}full}}%
\put(4950,9125){\makebox(0,0)[r]{\strut{}restricted}}%
\end{picture}%
\endgroup
 